\documentclass[a4paper,11pt]{article}
\pdfoutput=1
\usepackage{jheppub}

\usepackage{epsf}
\usepackage{youngtab}
\usepackage{multirow}

\newcommand{\beq}{\begin{eqnarray}}
\newcommand{\eeq}{\end{eqnarray}}

\newcommand{\SO}{\text{SO}}
\newcommand{\SU}{\text{SU}}
\newcommand{\U}{\text{U}}
\newcommand{\Sp}{\text{Sp}}

\newcommand{\refeq}[1]{Eq.(\ref{#1})}
\newcommand{\hhref}[2][]{\href{http://arxiv.org/abs/#2#1}{arXiv:#2}}
\newcommand{\met}{\not{\!\!{ E}}_{T}}
\newcommand{\mpt}{\not{\! { p}}_{T}}

\newcommand{\be}{\begin{equation}}
\newcommand{\ee}{\end{equation}}
\newcommand{\bea}{\begin{eqnarray}}
\newcommand{\eea}{\end{eqnarray}}

\makeatletter
\def\section{\@startsection {section}{1}{\z@}{-3.5ex plus -1ex minus
 -.2ex}{2.3ex plus .2ex}{\large\bf}}
\def\subsection{\@startsection{subsection}{2}{\z@}{-3.25ex plus -1ex
minus -.2ex}{1.5ex plus .2ex}{\normalsize\bf}}
\makeatother
\makeatletter

\@addtoreset{equation}{section}

\makeatother

\begin{document}

\title{Composite scalars at the LHC:\newline the Higgs, the Sextet and the Octet}
\date{\today}

\author[a,b]{Giacomo~Cacciapaglia,}
\author[a,b]{Haiying~Cai,}
\author[a,b,c]{Aldo~Deandrea,}
\author[d]{Thomas~Flacke,}
\author[d,e]{Seung~J.~Lee,}
\author[e]{Alberto~Parolini}


\affiliation[a]{Universit\'e de Lyon, France; Universit\'e Lyon 1, Villeurbanne, France}
\affiliation[b]{Institut de Physique Nucl\'eaire de Lyon, CNRS/IN2P3, UMR5822,\\ F-69622 Villeurbanne Cedex, France}
\affiliation[c]{Institut Universitaire de France, 103 boulevard Saint-Michel, 75005 Paris, France}
\affiliation[d]{Department of Physics, Korea University, Seoul 136-713, Korea}
\affiliation[e]{School of Physics, Korea Institute for Advanced Study, Seoul 130-722, Korea}

\emailAdd{g.cacciapaglia@ipnl.in2p3.fr}
\emailAdd{hcai@ipnl.in2p3.fr}
\emailAdd{a.deandrea@ipnl.in2p3.fr}
\emailAdd{flacke@korea.ac.kr}
\emailAdd{sjjlee@korea.edu}
\emailAdd{parolini85@kias.re.kr}

\abstract{
We present a phenomenological theory of scalar particles that transform as a sextet and an octet of QCD interactions. 
These particles may arise as light bound states of a fundamental dynamics giving rise to a composite Higgs boson and partial compositeness for the top.
As a concrete example, we discuss an explicit UV completion based on the SU(4)/Sp(4) coset, where QCD colour is carried by additional fundamental fermions charged under the confining gauge group.
Top partners, as well as potentially even lighter coloured scalars, arise as bound states of the coloured fermions.
We study production and detection at LHC Run I and II of the octet and sextet, setting lower limits on masses and couplings to Standard Model particles using existing 8 TeV analyses. We finally explore prospects for the ongoing 13 TeV Run II: we focus on final states with two same sign leptons, that have the potential to discriminate the sextet. 
}

\maketitle

\setcounter{tocdepth}{3}

\section{Introduction}
The Higgs boson of the Standard Model (SM) and the associated mechanism for the electroweak symmetry breaking is a striking and remarkably 
successful description of the observed electroweak physics, starting from low energy data up to the high energy colliders such as LEP and the LHC.
However, the quest for a more fundamental description of the electroweak and strong interactions is still wide open as not only fundamental theoretical 
questions await an answer, but also unexplained phenomena such as the origin of Dark Matter and the baryon asymmetry in the Universe.
The two main avenues for physics beyond the SM lead to consider the Higgs boson either as a truly fundamental scalar particle as part 
of a larger fundamental sector (as in many supersymmetric or non-supersymmetric extensions of the SM), or as a composite state of a more 
fundamental underlying dynamics. 
The latter possibility is very intriguing as spontaneous symmetry breaking via confinement is a phenomenon which is observed in nature in many systems, 
notably in Quantum Chromo-Dynamics (QCD) in a relativistic system setup, as well as in non-relativistic cases, for example in condensed matter systems. 
Moreover, from the theoretical side, asymptotically free and confining theories have a special 
status as these theories are potentially well defined at all scales. This second option is therefore theoretically appealing and was investigated in the 
past both at the effective and at the fundamental level. The earliest attempts in these directions were technicolor theories \cite{Weinberg:1975gm, Weinberg:1979bn, Susskind:1978ms}, which provided a simple testing ground for these ideas based on a scaled-up version of the QCD dynamics. It was soon clear that 
these first ideas needed to be implemented in a different way in order to allow for a light scalar boson. One key idea was to extend the global flavour 
symmetries of the underlying fermionic sector to generate a Higgs-like state as one of the pseudo-Nambu Goldstone Bosons (pNGB) of the theory, 
in a way similar to pions in QCD. In this way a composite scalar can be naturally and parametrically lighter than other composite particles in the 
theory \cite{Kaplan:1983fs,Kaplan:1983sm,Georgi:1984ef,Banks:1984gj,Georgi:1984af,Dugan:1984hq}. 
More recently, models based on a warped extra dimensional background have been proposed and studied~\cite{Contino:2003ve,Agashe:2004rs,Agashe:2005dk}, while being conjectured to be dual descriptions of four-dimensional conformal field theories (holographic Higgs)~\cite{Luty:2004ye}.
The experimental discovery of a light Higgs boson at the LHC 
has further motivated the detailed study of models of composite pNGB Higgs, based on the effective chiral Lagrangian approach 
\cite{Manohar:2006gz,Giudice:2007fh,Grinstein:2007iv,Barbieri:2007bh,Panico:2011pw,DeCurtis:2011yx,Alonso:2012px,Contino:2013kra}, mainly based on the minimal case that only contains a Higgs-like scalar as a light state.
Chiral Lagrangians depend on the choice of the spontaneous global symmetry breaking pattern and fermion representations, but even if there are 
many possible choices for phenomenologically viable effective Lagrangians~\cite{Katz:2005au,Gripaios:2009pe,Mrazek:2011iu}, it is not guaranteed that a fundamental UV description associated to a 
particular choice is possible in terms of a fundamental fermionic realisation. Therefore the effective chiral Lagrangian description of models of 
composite pNGB Higgs (usually called ``composite Higgs models'') do not provide a detailed understanding of the underlying physics. This is not a 
problem when trying to parameterise the Higgs sector, but one has to keep in mind that  assumptions of what is the light sector of the 
theory are implicit in such a description. In particular the standard assumption of these models is that apart from the Higgs boson the next light states 
are fermionic top partners~\cite{Contino:2006qr,Anastasiou:2009rv}. 
One helpful approach to resolve this impasse is to consider a UV completion based on fundamental interactions and fermions~\cite{Cacciapaglia:2014uja}, so as to complement and support the effective theory with a Fundamental Composite Dynamics (FCD) description. An alternative approach would be to rely on the existence of a conformal field theory in the UV, as described in the holographic approach.
A number of recent papers have been dedicated to the FCD approach, providing a detailed description ranging from 
model building~\cite{Cacciapaglia:2014uja,Ryttov:2008xe,Galloway:2010bp,Barnard:2013zea,Ferretti:2013kya,Ferretti:2014qta,Vecchi:2015fma}, lattice calculations~\cite{Hietanen:2014xca}, up to an overview of the phenomenology of the scalar and vector sectors 
\cite{Arbey:2015exa}. More in general Ref.s \cite{Bellazzini:2014yua,Panico:2015jxa} offer extensive and up to date reviews on model building efforts in the context of a composite Higgs. 

In this paper we focus on a set of possibly light states that are usually not included in the chiral Lagrangian for composite models with top partners: scalar mesons that carry colour. Such states should actually be expected, as the FCD description necessarily couples to QCD colour if coloured fermionic resonances are expected to appear in the low energy regime: an effective theory of scalar leptoquarks in composite Higgs models with partial compositeness is presented in \cite{Gripaios:2009dq}.
We will focus on a specific model proposed in~\cite{Barnard:2013zea}, which relies on the coset SU(4)/Sp(4) to realize the pNGB Higgs.
The SU(4)/Sp(4) coset is usually considered as the next to minimal ``composite Higgs model'' in the effective Lagrangian approach~\cite{Gripaios:2009pe}. However this is a minimal choice when considering a description in terms of bound states of fermions~\cite{Cacciapaglia:2014uja,Galloway:2010bp}. In this model the breaking is generated by an 
antisymmetric 6-dimensional representation (with respect to the global flavour symmetry) and the coset contains 5 Goldstone bosons. In terms of the 
custodial SO(4) subgroup of the residual symmetry group, the Goldstone bosons decompose into a $({\bf 2},{\bf 2}) + ({\bf 1},{\bf 1})$, which allows to obtain a pNGB Higgs. 
This case is also a simple example, which illustrates most of the  important features of UV descriptions for FCD in the electroweak sector. 
In Ref.~\cite{Barnard:2013zea}, top partners, i.e. coloured fermionic bound states, are obtained by adding 6 additional Weyl fermions and thus enlarging the global symmetry of the model with an extra SU(6) which embeds the SU(3)$_c$ of QCD. At this point we notice that the FCD provides a genuine UV completion for the electroweak sector but it does not encompass a dynamical explanation for the couplings of the top partners to the top: while values for these mixings reproducing the correct top mass are certainly allowed we do not investigate under which conditions they are obtained. The only pragmatic assumption we make is that the scale at which the four fermion interactions responsible for the mixing are generated is at least larger than the natural cut-off of the effective field theory, $4 \pi f$ where $f$ is the scale of the condensate. For this case to happen, the operator mixing to the top needs to have large anomalous dimensions: having a detailed model for the underlying theory, in principle, allows to compute such anomalous dimensions (on the Lattice) and investigate the UV origin of the four fermion interactions. In this paper, however, we limit ourselves to an effective theory, where we take fully into account the symmetries of the underlying theory. We will focus, in particular, on the presence of coloured scalar resonances in the low energy theory. The breaking of the global SU(6) by an explicit mass term and the strong dynamics then implies the presence of 20 pNGBs in the spectrum, which decompose to a real colour octet and a complex charged sextet.
After briefly discussing the connection between the masses of the coloured pNGBs and the masses of the top partners (which are needed to be light for naturalness arguments, and to generate the correct top mass), we consider the phenomenology of such states at the LHC.
We consider in particular the case where they are the lightest composite coloured states in the spectrum, and then study the bounds from the LHC Run I data and the prospects to distinguish the presence of a sextet versus the octet at the LHC Run II.
The analysis is done by using an effective description of their interactions, thus providing a model independent determination of their phenomenology.

The paper is organised as follows: in section \ref{sec:model} we discuss the structure of the model and its bound states. 
Section \ref{sec:bound} is dedicated to constructing the effective description stemming from the fundamental model of a peculiar exotic scalar sextet 
which has potentially interesting collider signatures. Section \ref{sec:LHC} explores the signatures 
which are expected at the LHC for the exotic scalar sextets and octets present in the model and how to distinguish them in the LHC set-up. 
We discuss the conclusions which can be drawn on this class of models and the preliminary exploration of their phenomenology at the LHC in 
section \ref{sec:conclusion}.

\section{The model: SU(4)/Sp(4) coset based on $G_{\rm FCD} =$ Sp($2 N_c$)}
\label{sec:model}
In order to explore the composite Higgs idea from the fundamental perspective of underlying constituent fermionic states, we consider 
a minimal model which was already outlined in the literature and partially explored in some of its phenomenological aspects.
The model contains four Weyl techni-fermions $Q_i$ in the spinorial representation of Sp($2N_c$): the minimal choice is SU(2) with fermions in the fundamental~\cite{Ryttov:2008xe,Galloway:2010bp}.
It has been shown on the lattice~\cite{Hietanen:2014xca,Lewis:2011zb,Hietanen:2013fya} that the global symmetry SU(4), acting on the $Q$'s, is then dynamically broken to Sp(4), leading to five Goldstone bosons: three remain exact Goldstones and are eaten by the $W$ and $Z$, one plays the role of the Higgs, and the fifth is a gauge singlet.
Additional six techni-fermions $\chi_j$ in the 2-index anti-symmetric representation are needed to generate coloured fermionic bound states~\cite{Barnard:2013zea,Ferretti:2013kya}, i.e. the top partners necessary to realize partial compositeness.
This requirement imposes that the minimal FCD group is Sp(4), while larger FCD groups seem to be disfavoured: it has been shown that this model can satisfy electroweak precision tests~\cite{Arbey:2015exa} when the top partner sector is not included, and that it may be close to the conformal window once the $\chi$'s are added~\cite{Sannino:2009aw}.
Larger $N_c$ will lead to larger electroweak corrections, and to models that are deeper in the conformal window. Also vector and fermion resonances induce non negligible effects on precision parameters: in general they are model dependent and they can be large, potentially dangerous. Studies on models based on the minimal coset $\SO(5)/\SO(4)$ show that there still exist available regions of parameter space, and therefore we expect the same to hold in the present case.
The techni-fermions are charged under the FCD and SM gauge symmetries Sp(2$N_c$)$\times$SU(3)$_c \times$ SU(2)$_L \times$ U(1)$_Y$ as reported in Table~\ref{tab:fund_field_content}.
\begin{table}[tb]
\begin{center}
\begin{tabular}{|c|c|c|c|c||c|c|c|}
\hline
& $\Sp(2N_c)$&${\SU(3)}_c$&${\SU(2)}_L$&${\U(1)}_Y$ & SU(4) & SU(6) & U(1) \\
\hline
$ \begin{array}{c} Q_1 \\ Q_2 \end{array} $&${\tiny{\yng(1)}}$&$\bf 1$&$\bf 2$&$0$& \multirow{3}{*}{\bf 4} & \multirow{3}{*}{\bf 1} & \multirow{3}{*}{$-3(N_c-1)q_\chi$}\\
\cline{1-5}
$Q_3$&${\tiny{\yng(1)}}$&$\bf 1$&$\bf 1$&$1/2$ & & &\\
\cline{1-5}
$Q_4$&${\tiny{\yng(1)}}$&$\bf 1$&$\bf 1$&$-1/2$ & & & \\
\hline
$ \begin{array}{c} \chi_1 \\ \chi_2 \\ \chi_3\end{array} $&${\tiny{\yng(1,1)}}$&$\bf 3$&$\bf 1$&$x$ & \multirow{4}{*}{\bf 1} & \multirow{4}{*}{\bf 6} & \multirow{4}{*}{$q_\chi$}\\
\cline{1-5}
$ \begin{array}{c} \chi_4 \\ \chi_5 \\ \chi_6\end{array} $&${\tiny{\yng(1,1)}}$&$\bf \bar{3}$&$\bf 1$&$-x$ & & &\\
\hline
\end{tabular} 
\caption{Field content of the microscopic fundamental theory and property transformation under the gauged symmetry group Sp(2$N_c$)$\times$SU(3)$_c \times$ SU(2)$_L \times$ U(1)$_Y$, and under the global symmetries SU(4)$\times$SU(6)$\times$U(1).}
\label{tab:fund_field_content}
\end{center}
\end{table}

With respect to the FCD, namely neglecting the SM gauging, the techni-fermions transform under a global symmetry SU(4)$\times$SU(6)$\times$U(1), with SU(2)$_L \subset$ SU(4), SU(3)$_c \subset$ SU(6) and $\U{(1)}_Y \subset\SU(4)\times\SU(6)$. The SM hypercharge is obtained gauging a combination of a ${\U(1)}_x \subset$ SU(6), and a ${\U(1)}$ included in the custodial $\SO(4)\subset\SU(4)$. The hypercharge $x$ can be determined by the choice of which top partners couple with the top quarks. The global $\U(1)$ is the linear combination of $Q$ and $\chi$ number which is not anomalous and whose charges are defined by $q_Q=-3(N_c-1)q_\chi$.

Here we will not study the details of the dynamics leading to the breaking of the global symmetries, as a discussion in terms of a Nambu-Jona-Lasinio model can be found in~\cite{Barnard:2013zea}: we will simply assume that both $QQ$ and $\chi \chi$ form a condensate.
Note that the value of the fermion bilinear operator on the vacuum will depend both on the spontaneous breaking induced by the strong dynamics and on explicit breaking terms, like the masses of the fundamental fermions.
Once the gauge $\Sp(2 N_c)$ interactions condense, two scalar bound states can form:
\begin{enumerate}
\item $\langle Q Q \rangle$, transforming as $({\bf 6}, {\bf 1}, 2 q_Q )$ under the flavour SU(4)$\times$SU(6)$\times$U(1) global symmetries. This object is the one responsible for the breaking of the SU(4) symmetry, and therefore the EW symmetry. The breaking pattern is SU(4)$\to$ Sp(4). Note that this condensate will also break the global U(1).
\item $\langle \chi \chi \rangle$, transforming as $({\bf 1}, {\bf 21}, 2 q_\chi)$ under the flavour symmetry. A non-zero value for its condensate corresponds to the mass term added in \cite{Barnard:2013zea}, i.e. a VEV for $\langle \chi\chi \rangle$. Thus, the mass term for the $\chi$'s explicitly breaks SU(6)$\to$SO(6)$\sim$SU(4). Note that SO(6) contains a subgroup ${\SU(3)}_c\times{\U(1)}_X$. This condensate also breaks the global U(1).\footnote{Note that all $\chi$'s belong to the same representation of $G_{\rm FCD}$, thus there is a global SU(6) symmetry rather than 
SU(3)$\times$SU(3)~\cite{Barnard:2013zea}. The mass term can be expressed as a gauge-invariant combination of the $\chi$'s, which is invariant under a global 
SO(6).}
\end{enumerate}

\subsection{Bound States}

We can now classify all the bound states in terms of their transformation property under the global flavour symmetry SU(4)$\times$SU(6), and their 
unbroken subgroups Sp(4)$\times$SO(6) as shown in Table~\ref{tab:bdstates}.
\begin{table}[tb]
\begin{center}
\begin{tabular}{|cc|c|c|c|}
\hline
  & spin & SU(4)$\times$SU(6) & Sp(4)$\times$SO(6) & names \\
\hline \hline
$QQ$ & 0 & $(\bf{6}, \bf{1})$ &   $(\bf{1}, \bf{1})$ & $\sigma$ \\
          &     &                       &   $(\bf{5}, \bf{1})$ & $\pi$ \\
\hline
$\chi \chi$ & 0 & $(\bf{1}, \bf{21})$ & $(\bf{1}, \bf{1})$ & $\sigma_c$ \\
                 &    &                          & $(\bf{1}, \bf{20})$ & $\pi_c$ \\
\hline \hline
$\chi Q Q$ & 1/2 & $(\bf{6}, \bf{6})$ & $(\bf{1}, \bf{6})$ & $\psi_1^1$ \\
                  &       &                       & $(\bf{5}, \bf{6})$ & $\psi_1^5$ \\
\hline
$\chi \bar{Q} \bar{Q}$ & 1/2 & $(\bf{6}, \bf{6})$ & $(\bf{1}, \bf{6})$ & $\psi_2^1$ \\
                  &       &                       & $(\bf{5}, \bf{6})$ & $\psi_2^5$ \\
\hline
$Q \bar{\chi} \bar{Q}$ & 1/2 & $(\bf{1}, \bf{\bar{6}})$ & $(\bf{1}, \bf{6})$ & $\psi_3$ \\
\hline
$Q \bar{\chi} \bar{Q}$ & 1/2 & $(\bf{15}, \bf{\bar{6}})$ & $(\bf{5}, \bf{6})$ & $\psi_4^5$ \\
                               &        &                                  & $(\bf{10}, \bf{6})$ & $\psi_4^{10}$ \\
\hline \hline
$\bar{Q} \sigma^\mu Q$ & 1 & $(\bf{15}, \bf{1})$ & $(\bf{5}, \bf{1})$ & $a$ \\
                                       &    &                          & $(\bf{10}, \bf{1})$ & $\rho$ \\
\hline
$\bar{\chi} \sigma^\mu \chi$ & 1 & $(\bf{1}, \bf{35})$ & $(\bf{1}, \bf{20})$ & $a_c$ \\
                                       &    &                          & $(\bf{1}, \bf{15})$ & $\rho_c$ \\
\hline
\end{tabular} 
\caption{Bound states of the model with spin and group properties with respect to the global flavour group and the unbroken subgroups.}
\label{tab:bdstates}
\end{center}
\end{table}
The SU(4) flavour symmetry is broken by the $QQ$ condensate, which transforms as a 2-index anti-symmetric in flavour, i.e. as a $\bf 6$ of SU(4). The breaking pattern is:
\beq
\mbox{SU(4)} \to \mbox{Sp(4)}\,, \quad \mbox{with}\;\; 15 - 10 = 5\;\; \mbox{pseudo-Goldstone bosons ($\pi$).}
\eeq
For the SU(6) flavour symmetry, the condensate $\chi \chi$, transforming as a 2-index symmetric of SU(6), i.e. $\bf 21$, breaks
\beq
\mbox{SU(6)} \to \mbox{SO(6)}\,, \quad \mbox{with}\;\; 35 - 15 = 20\;\; \mbox{pseudo-Goldstone bosons ($\pi_c$).}
\eeq
Note that the number of pseudo-Goldstones matches the dimensions of the $\pi$ and $\pi_c$ scalars.
The 20 degrees of freedom in $\pi_c$ correspond to the mesons $R$, $P$ and $\tilde{P}$ in~\cite{Barnard:2013zea}, while the singlet $S$ is the analog of $\sigma_c$.
Note also that the breaking of the global U(1) leads to an additional light singlet, which is a combination of $\sigma$ and $\sigma_c$, while the orthogonal combination, being associated to the anomalous U(1), will develop a large mass (similarly to the $\eta'$ in QCD).

The detailed spectrum of the low energy theory can only be studied numerically on the lattice. Naive expectation may lead us to guess that the pseudo-Goldstones $\pi$ and $\pi_c$ are parametrically lighter that the other states, while the scalar singlet, the spin-1/2 states and the vectors pick up a mass at the order of the condensation scales. As the SU(6) symmetry is broken explicitly by a mass term, the coloured pNGBs will be expected to have a mass of this order, however the $\chi$ mass will also contribute to a mass term for the spin-1/2 top partners.
In fact, we can see from the table that it is not possible to write an SU(6)-invariant mass term for the composite fermions~\footnote{We use Weyl fermion notation here, so a mass term will always be written as $\psi \psi'$, thus transforming as a {\bf 6}$\otimes${\bf 6}$= ${\bf 21}$\oplus${\bf 15} of SU(6).}: we can therefore guess that the mass of the fermions will receive contributions from the dynamical and explicit SU(6) breaking, while the pseudo-Goldstones $\pi_c$ will only receive a contribution from the explicit breaking.
On the other hand, the spin-1 states can have an SU(6) invariant mass.
The expected hierarchy in the spectrum of coloured states is thus that the spin-1 are the heaviest states, while the $\pi_c$ are the lightest.

This naive scenario can be altered when considering that the model is close to the conformal window, thus large anomalous dimensions may be generated for some of its composite operators. 
In particular, one may wish for a large anomalous dimension for the top partners, that brings down their mass close to 1 TeV in order to generate the correct top mass via partial compositeness.
One qualitative argument is based on the observation that all top partners contain a $QQ$ pair (in terms of the FCD, $\bar{Q} Q$ is equivalent to $QQ$), which may be more tightly bound than the additional $\chi$: in other words, we can assume for simplicity that the top partners are composed of a tightly bound $QQ$ pair connected to a $\chi$~\cite{Barnard:2013zea}.
However, the $QQ$ inside the top partners transforms as a 2-index antisymmetric of $\Sp(2N_c)$, while the $QQ$ pair that condenses is a singlet. We can then use the Maximally Attractive Channel (MAC) \cite{Raby:1979my} reasoning: the force between two fermions in representations $r_1$ and $r_2$ of the strong 
dynamics which form a bound state in the $r_{12}$ representation is proportional to 
\begin{equation}
A = C_2 (r_1) + C_2 (r_2) - C_2 (r_{12})\,,
\end{equation}
thus the channel with larger $A$ is more tightly bound. When combining $QQ$, the singlet channel (condensate) with $C_2(r_{12}) = 0$ is more 
attractive than the other one ($QQ$ in the composite fermion) which has $C_2 (r_{12}) > 0$. From this argument follows that the $QQ$ pair inside 
the fermion is bound in a weaker way than the $QQ$ pair in the condensate. 
One may therefore naively expect that the anomalous dimension of the $QQ$ condensate is larger that the one of the top partners. 
Furthermore, all top partners in Table~\ref{tab:bdstates} share the same structure in terms of the FCD as $Q$ is in a pseudo-real and $\chi$ in a real representation, therefore the anomalous dimensions of the top partners should be very close to each other.

From the above arguments it follows that one may well expect that the dynamical mass of the top partners and of the scalars $\pi_c$ are in the same ballpark.
In the following, we will focus on this scenario, which contrasts with the choice made in \cite{Barnard:2013zea} where the 6-plets $\psi_{1,2}$ made of $\chi Q Q$ bound states were chosen lighter than the other composite states, and the mesons $\pi_c$ are considered much heavier. 
Note that the choice in \cite{Barnard:2013zea} corresponds to the standard assumptions behind the model building of composite Higgs effective theories.

\subsubsection{Mesons}

The theory stemming from the structure we detailed above contains different bound state particles, which can be studied according to their 
statistics and their quantum numbers. A first class of particles are those analogous to the mesons which are found as bound states in strong interactions. 
In the limit where the condensates are aligned in a direction that does not break the SM gauge symmetries, we can use them to classify the various states.
The mesons decompose as:
\beq\label{eq:decomposition scalars}
\pi &=& ({\bf{1}}, {\bf{2}}, {\bf 2})_0 \oplus ({\bf1},{\bf1},{\bf1})_0\,, \nonumber \\
\pi_c &=& ({\bf 8},{\bf1},{\bf1})_0 \oplus ({\bf6},{\bf1},{\bf1})_{2x} \oplus ({\bf\bar{ 6}},{\bf1},{\bf1})_{-2x}\,,  \nonumber\\
\sigma (\sigma_c) &=& ({\bf1},{\bf1},{\bf1})_0\,; \nonumber
\eeq
where the representations under SU(3)$_c \times$ SU(2)$_L\times$SU(2)$_R$ are indicated by the numbers in parenthesis, and the subscript corresponds to the 
charge under U(1)$_x$. We keep explicit track of the custodial symmetry embedded in the model, while the hypercharge U(1)$_Y$ is the gauged subgroup of SU(2)$_R \times$U(1)$_x$.

The 5-plet $\pi$ contains a bi-doublet that will play the role of the Brout-Englert-Higgs doublet, plus a singlet ($\eta$): once the condensate is misaligned in a direction that breaks the EW symmetry, the quantum numbers of the various states will not coincide any more with the quoted ones.
Anyway, one can still think of the Higgs candidate as a leftover of the bi-doublet, while the singlet $\eta$ will acquire some couplings to the SM gauge bosons.
The phenomenology of the Higgs candidate and singlet states, in a model without top partners, has been studied in detail in~\cite{Arbey:2015exa}.
The coloured scalars arise form the breaking of the SU(6) symmetry, necessary to give mass to the top partners without breaking any of the SM gauge symmetries: they will therefore play no role in the Higgs physics, however they will affect the phenomenology of the top partner sector, and they will be the focus of the present work.
Other two scalars, $\sigma$ and $\sigma_c$, are the pseudo Goldstone bosons of the ${\U(1)}_Q$ and ${\U(1)}_\chi$ associated to the $Q$ and $\chi$ numbers, and they are SM singlets; a combination of the two gets mass via $\Sp(2N_c)$ instanton effects, corresponding to the anomalous global U(1). The orthogonal combination remains a pNGB, and receives a mass from the explicit breaking terms, like the masses of the fundamental fermions $\chi$ and $Q$.

\subsubsection*{Coloured pNGB masses}

We now discuss more specifically the masses of the coloured pNGBs, which will be the object of the more detailed phenomenological analysis. 
The embedding of QCD SU(3)$_c$ in the global SU(6) as in Table~\ref{tab:fund_field_content} allows to write a mass term for $\chi$ as
\beq
\mathcal{L}_{\rm FCD} \supset m_{\chi}\; \chi^T \cdot \left( \begin{array}{cc}
0 & 1_{3\times 3} \\
1_{3\times 3} & 0
\end{array} \right) \cdot \chi + h.c.\,.
\eeq
The SU(3) preserving vacuum is therefore aligned with the mass matrix
\beq
\Sigma_{\chi \chi} = \left( \begin{array}{cc}
0 & 1_{3\times 3} \\
1_{3\times 3} & 0
\end{array} \right)\,,
\eeq
and it breaks SU(6) $\to$ SO(6), as expected.
The unbroken gauged subgroup SU(3)$_c \times$ U(1)$_x$ is thus
\beq
S^a = \frac{1}{\sqrt{2}} \left( \begin{array}{cc}
\lambda^a & 0 \\
0 & - {\lambda^a}^T
\end{array} \right)\,, \qquad X = x  \left( \begin{array}{cc}
1 & 0 \\
0 & - 1
\end{array} \right)\,,
\eeq
where $\lambda^a$ are the Gell-Mann matrices (generators of SU(3)$_c$).
The coloured pions can be written as
\beq \label{eq:U6}
U_6 = e^{i \Pi/f_6}\,, \qquad \Pi = \frac{1}{\sqrt{2}} \left( \begin{array}{cc}
\pi_8^a \lambda^a & \pi_6 \\
\pi_6^c & \pi_8^a {\lambda^a}^T
\end{array} \right)\,,
\eeq
where $\pi_6$ is a 3$\times$3 symmetric matrix containing a complex colour sextet, and $f_6$ is the decay constant which is in general different with the decay constant $f$ appearing in the $QQ$ condensate. Note that the mass of the $W$ and $Z$, and the scale of the electroweak symmetry breaking, are related to $f$.

Masses for the pions are generated by the explicit breaking terms of the global symmetry SU(6): the $\chi$ mass, gauge interactions and the couplings to the top.
The contribution of the $\chi$ mass can be expressed as:
\beq
\mathcal{L}_{\rm EFT} \supset
 C_\chi m_\chi f_6^3\, \mbox{Tr} \left[ \Sigma_{\chi\chi} \cdot U_6 \cdot \Sigma_{\chi\chi}\right] + h.c. \to M_\pi^2 \left( \frac{1}{2} \pi_8^2 + \pi_6^c \pi_6 \right) + \dots
\eeq
where $M_\pi^2 \sim C_\chi m_\chi f_6$, $C_\chi$ being a numerical factor.
Gauge interactions contribute at one loop via the QCD
\beq
\mathcal{L}_{\rm EFT} \supset g_s^2 C_g f_6^4\, \mbox{Tr} \left[ S^a \cdot U_6 \cdot \Sigma_{\chi\chi} \cdot (S^a\cdot U_6 \cdot \Sigma_{\chi\chi})^\ast \right] \sim C_g f^2_6 g_s^2 \, \left( \frac{3}{4} \frac{\pi_8^2}{2} + \frac{5}{6} \pi_6^c \pi_6\right) + \dots
\eeq
and U(1)$_x$
\beq
\mathcal{L}_{\rm EFT} \supset
{g'}^2 C_g f_6^4 \, \mbox{Tr} \left[ Y \cdot U_6 \cdot \Sigma_{\chi\chi} \cdot (Y\cdot U_6 \cdot \Sigma_{\chi\chi})^\ast \right] \sim C_g f^2_6 {g'}^2 2 x^2 \, \pi_6^c \pi_6 + \dots
\eeq
contributions.
The coefficients in front of the QCD loop can be easily understood in terms of the Casimir of the two representations: $C_2({\bf8}) = 3$ and $C_2({\bf6}) = 10/3$.
Both contributions are expected to be positive, i.e. $C_\chi>0$ and $C_g>0$, else they would induce a VEV that breaks gauge interactions themselves.
The contribution of top loops can be estimated by a spurionic analysis of the top couplings to the coloured pNGBs: all the composite baryons transform as the fundamental (or anti-fundamental) representation of SU(6), thus the elementary quark fields should be embedded in an anti-fundamental (or fundamental) in order for linear couplings to be written.
We can thus associate the elementary fields to two spurions in, e.g., the ${\bf \bar{6}}$ of SU(6):
\beq
P_L^a = ( 0, \delta^{i,a} )^T\,, \quad P_R^{\bar{a}} = (\delta^{j, \bar{a}}, 0)^T\,,
\eeq
where $a$ and $\bar{a}$ are indices of the QCD colour, and $i = 1,2,3$ and $j = 4,5,6$ run over the SU(6) indices.
We can now build an operator, invariant under the stability group SO(6), as
\beq
\mathcal{O} = (U^\dagger_6 \cdot P_X)^T \cdot \Sigma_{\chi\chi} \cdot U_8^\dagger \cdot P_Y\,, \quad X,Y = L, R\,.
\eeq
The projectors $P_X$ select a $3\times3$ sub-block of $\mathcal{O}$: the off-diagonal block, transforming as a $1 \oplus 8$ of QCD colour, for L--R, the first diagonal block, transforming as a $\bar{6}$, for R--R, and the second diagonal block, transforming as a $6$, for L--L.
The operator above can be used to construct effective couplings of the pNGBs to the tops, as follows:
\beq
& c_{LR}\, v \; \left( 1 - \frac{\sqrt{2} i}{f_6} \pi_8 - \frac{1}{f_6^2} \left( \pi_6 \pi_6^c + \pi_8^2 \right) + \dots \right)\, t_L t_R^c\,, &  \nonumber \\
& c_{LL}\, \frac{v^2}{f} \left( - \frac{\sqrt{2} i}{f_6} \pi_6 - \frac{1}{f_6^2} \left( \pi_8 \pi_6 + \pi_6 \pi_8^T \right) + \dots \right) t^c_L t^c_L\,, & \label{eq:tcoup} \\
&c_{RR}\, f \left( - \frac{\sqrt{2} i}{f_6} \pi_6^c - \frac{1}{f_6^2} \left( \pi_8^T \pi_6^c + \pi_6^c \pi_8 \right) + \dots \right) t_R t_R\,, \nonumber &
\eeq
where the powers of $v$ come from the SU(4)/Sp(4) structure and derive from the transformation properties of the top bilinear under SU(2)$_L$ (we keep only the leading term in $v/f$), while the coefficients $c_{XY}$ are quadratic in the pre-Yukawas and their form depends on the specific representation of Sp(4) the top partners belong to. The above formula is very general, and it only depends on the SU(6) representation of the top partners.
The contribution to the pNGB masses can now be calculated by computing loops of the above operators: up to order $(v/f)^2$, we obtain
\beq
\delta m^2_{\pi_6} = - \frac{\Lambda^2}{8 \pi^2} \left( C_{RR} c_{RR}^2 \frac{f^2}{f_6^2} - C_{LR} c_{LR}^2 \frac{v^2}{f_6^2} + \mathcal{O} (v/f)^4 \right)\,,
\eeq
while no correction to the octet mass is generated. In the above formula, $C_{RR}$ and $C_{LR}$ are numerical $O (1)$ coefficients depending on the dynamics, and cut-off of the integral can be approximated by $\Lambda \sim 4 \pi f$. This mass correction is expected to be negative.

All in all, the masses of the pions can be written as:
\beq
m_{\pi_8}^2 &=& M_\pi^2 + C_g f_6^2 \left( \frac{3}{4} g_s^2 \right)\,, \\
m_{\pi_6}^2 &=& M_\pi^2 + C_g f_6^2 \left( \frac{5}{6} g_s^2 + 2 x^2 {g'}^2 \right) - C_{RR} f_6^2 c_{RR}^2 \frac{2 f^4}{f_6^4}\,.
\eeq
The mass hierarchy crucially depends on the size of the top loops:
\beq
m_{\pi_6}^2 - m_{\pi_8}^2 = - C_{RR} f_6^2 c_{RR}^2 \frac{2 f^4}{f_6^4} + C_g f_6^2 \left( \frac{1}{12} g_s^2 + 2 x^2 {g'}^2 \right)\,.
\eeq
If the top loop dominates, the sextet can be expected to be lighter than the octet Nevertheless this contribution can be made small in various ways: for instance, by reducing the coupling of the right-handed tops, or by generating a hierarchy $f < f_6$. The contribution of the gauge loops is more model independent, and it generates a numerically small mass splitting in favour of the octet.

\subsubsection{Baryons (top partners)}
\label{sec:baryons}
A similar decomposition can be obtained for the fermionic bound states of the theory, among which one can identify the top partners involved in the generation of the top mass:
\beq
\psi_{1,2}^{(1)} &=& ({\bf3}, {\bf1},{\bf1})_{x} \oplus ({\bf\bar{3}}, {\bf1},{\bf1})_{-x}\,, \nonumber \\
\psi_{1,2}^{(5)} &=& ({\bf3},{\bf1},{\bf1})_{x} \oplus ({\bf3},{\bf2},{\bf2})_{x} \oplus ({\bf\bar{3}},{\bf1},{\bf1})_{-x} \oplus ({\bf\bar{3}},{\bf2},{\bf2})_{-x}\,,  \nonumber \\
\psi_3 &=& ({\bf3}, {\bf1},{\bf1})_{x} \oplus ({\bf\bar{3}}, {\bf1},{\bf1})_{-x}\,,  \nonumber \\
\psi_4^{(5)} &=&  ({\bf3}, {\bf1},{\bf1})_{x} \oplus ({\bf3},{\bf2},{\bf2})_{x} \oplus ({\bf\bar{3}},{\bf1},{\bf1})_{-x} \oplus ({\bf\bar{3}},{\bf2},{\bf2})_{-x}\,,  \nonumber \\
\psi_4^{(10)} &=&  ({\bf3},{\bf2},{\bf2})_{x} \oplus ({\bf3},{\bf3},{\bf1})_{x} \oplus ({\bf3},{\bf1},{\bf3})_{x} \oplus  ({\bf\bar{3}},{\bf2},{\bf2})_{-x}  \oplus ({\bf\bar{3}},{\bf3},{\bf1})_{-x} \oplus ({\bf\bar{3}},{\bf1},{\bf3})_{-x}\,.  \nonumber 
\eeq
Note that, if we choose $x = 2/3$, the fermion spectrum contains the following states:
\begin{itemize}
\item[-] 4 $t_L$ candidates contained in the bi-doublets $({\bf3},{\bf2},{\bf2})_{2/3}$ which can mix with the elementary left-handed doublets: they are contained in $\psi_{1,2}^{(5)}$, $\psi_4^{(5)}$ and $\psi^{(10)}$;
\item[-] 7 $t_R$ candidates: 6 in the form of $({\bf\bar{3}},{\bf1},{\bf1})_{-2/3}$, present in $\psi_{1,2}^{(1)}$, $\psi_{1,2}^{(5)}$,  $\psi_3$ and $\psi_4^{(5)}$, and one contained in the SU(2)$_R$ triplet $({\bf\bar{3}},{\bf1},{\bf3})_{-2/3}$ in $\psi_4^{(10)}$;
\item[-] 1 $b_R$ candidate contained in the SU(2)$_R$ triplet $({\bf\bar{3}},{\bf1},{\bf3})_{-2/3}$ in $\psi_4^{(4)}$.
\end{itemize}
From this simple counting we can see that, while it would be possible in principle to give mass to all up-type quarks via partial compositeness, only one bottom state can receive its mass via this mechanism.
In order to generate a sufficient number of partners for all SM quarks (and leptons), without changing the Higgs sector, the number of $\chi$ should be increased: besides the generation of many more fermionic and scalar states, the model would be more in danger of falling inside the conformal window. We will therefore consider here only the minimal case.
A general study of the possible couplings in the fermionic sector is beyond the scope of this work. A discussion of the possible role of various representations of the SU(4) flavour symmetry, for the purpose of giving mass to the top, can be found in~\cite{Gripaios:2009pe}. We assume that some among these baryonic resonances couple linearly to left- and right-handed top, realizing the partial compositeness paradigm. These mixings are four fermion interactions in terms of the constituent fermions, generated at a certain scale. Notice that for this mechanism to work we need some physical process generating these interactions and proper values for couplings and anomalous dimensions so to reproduce the correct top mass: none of these is automatically in place in the present model. We put ourselves in the working assumption that suitable interactions are generated by a UV dynamics.

To counter the absence of partners for the light quarks and leptons, their masses can be generated assuming couplings of quark bi-linears with operators interpolating the Higgs field: in the UV they originate from four fermion interactions.
If such terms are only needed to generate a mass as large as the charm in the up sector, and the strange in the down sector, while both top and bottom are partially composite, then it can be shown that no flavour symmetries are necessary to protect the model against flavour bounds \cite{Cacciapaglia:2015dsa}: this conclusion is general, and it does not depend on the representation of the multiplet the top (and bottom) partners belong to.

The model contains vector-like quark partners that arise as $\SU(2)_L$ and $\SU(2)_R$ singlets, doublets and triplets, with electromagnetic charges (when choosing $x=2/3$) ranging from $-1/3$ to $5/3$.
Vector-like quarks with such properties and coupling to third generation quarks have been extensively studied at the LHC, and after Run I their masses are constrained to be heavier than $\sim 700 - 900\sim\mbox{GeV}$ (depending on the branching ratios into third generation quark and $h,W,Z$). In the same modes, a sensitivity up to $\sim 1.4~\mbox{TeV}$ \cite{Matsedonskyi:2014mna} (or even higher \cite{Backovic:2014uma}) are to be expected for Run II. As in the model under consideration there are several top partners with the same electro magnetic charge, their signal cross sections for the respective final states add, such that bounds on the resonance mass can be expected to be larger. 

\subsubsection{Vectors ($\rho$ and $a$)}
Spin-1 bound states are a typical prediction of a large class of composite models and are also considered in effective chiral Lagrangian type
models describing a composite or strongly interacting electroweak sector~\cite{Casalbuoni:1985kq}. In the present model we have the following states:
\beq
\rho &=&  ({\bf1},{\bf2},{\bf2})_{0} \oplus ({\bf1},{\bf3},{\bf1})_{0} \oplus ({\bf1},{\bf1},{\bf3})_{0}\, , \nonumber \\
a & = &   ({\bf1},{\bf1},{\bf1})_{0} \oplus ({\bf1},{\bf2},{\bf2})_{0} \,,  \nonumber \\
\rho_c &=& ({\bf1},{\bf1},{\bf1})_0 \oplus  ({\bf8},{\bf1},{\bf1})_0 \oplus ({\bf\bar{3}},{\bf1},{\bf1})_{2x} \oplus ({\bf3},{\bf1},{\bf1})_{-2x}\,,  \nonumber \\
a_c &=& ({\bf8},{\bf1},{\bf1})_0 \oplus ({\bf6},{\bf1},{\bf1})_{2x} \oplus ({\bf\bar{6}},{\bf1},{\bf1})_{-2x}\,.  \nonumber 
\eeq
The $\rho$ and $\rho_c$ states correspond to ``vector'' resonances in QCD, transforming as the adjoint representation of the unbroken global symmetries, while the $a$ and $a_c$ correspond to the ``axial'' resonances, associated with the broken generators and thus transforming in the same representation as the pseudo-Goldstone mesons.
 Note that the $\rho$'s contain a triplet of SU(2)$_L$ and a triplet of SU(2)$_R$, which can mix with the $W$ and $Z$. These states can be constrained by perturbative unitarity and LHC direct search, like in minimal models~\cite{Contino:2011np, Bellazzini:2012tv, Cai:2013ira,  Pappadopulo:2014qza, Greco:2014aza}. The $a$'s contain a $4$-plet  axial vector resonance.  Its phenomenology related to Higgs decay is studied in  a minimal $\SO(5)/\SO(4)$ model~\cite{Cai:2013}. $\rho_c$ contains a ``KK gluon'', i.e. a colour octet { \it c.f. e.g.} \cite{Chen:2014haa} for a recent study.  This model also contains  a  colour sextet  vector in the coset space,  its  collider simulation is explored in \cite{Han:2010rf, Zhang:2010kr}. The phenomenology of such states have been widely studied in the literature, and several searches from both ATLAS and CMS can be used to impose strict bounds on their masses, for example the recent experimental studies concerning same sign leptonic final states or multi-jet final states \cite{atlasSS,atlasSS2,atlastt,atlastt2,cmsSS,cmstt,cmstt2,cmstt3} which can be used to extract approximate bounds recasting these analyses. For the model presented here, these bounds apply only if the $\rho_c$ states do not cascade decay through either top partners~\cite{Bini:2011zb}, or pseudo-goldstone bosons. Furthermore, the model under consideration contains additional states, like the ``axial'' colour sextet $a_c$, the colour triplet  vectors, or the weak doublets, whose phenomenology deserves further studies.

Note, however, that spin-1 vectors are expected to be more massive than the meson bound states in the present class of models: this has been shown in the minimal model without top partners on the lattice~\cite{Hietanen:2014xca,Hietanen:2013fya}, where the spin-1 resonances appear at a scale above 3~TeV. In the present model, the near conformal dynamics may bring down the masses, however we would naively expect the hierarchy between masses, sketched above, to be preserved.

\subsection{Couplings of the coloured pNGBs} \label{sec:couplings}

The couplings of the coloured pions to fermions is relevant for the phenomenological study we perform in the following sections and can be obtained as follows. 
Consider generic composite top partner:
\beq
\Psi = \left( \begin{array}{c} 
\psi \\ \eta \end{array} \right)
\eeq
transforming as a $({\bf R},{\bf 6})$ of the unbroken Sp(4)$\times$SO(6) stability group.
Here, $\psi$ is a colour-triplet, while $\eta$ is an anti-triplet, and both are left-handed Weyl spinors. This state can be embedded into an object transforming linearly under the full SU(4)$\times$SU(6) group by inserting appropriate pion matrices:
\beq
\tilde{\Psi} = U_6 \cdot \left( \begin{array}{c} U_4^{(R)}  \psi \\ U_4^{(R)} \eta \end{array} \right)\,,
\eeq
where $U_4^{(R)}$ is the pion matrix of SU(4)/Sp(4) in the representation $\bf R$ and $U_6$ is given in \refeq{eq:U6}.
The linear mixing term of the left handed top with top partners can then be written as:
\beq
y_L  f \, (0, \xi_L) \cdot \tilde{\Psi} = y_L f\,  \left( 1 - \frac{i}{\sqrt{2} f_6} \pi_8^a {\lambda^a}^T\right) \xi_L U_4^{(R)} \eta - y_L f \,\frac{i}{\sqrt{2} f_6}  \pi_6^c \xi_L U_4^{(R)} \psi + \dots \label{eq:leftpreY}
\eeq
where $\xi_L$ is the spurion containing the left-handed quarks in the representation $\bf \bar{R}$ of SU(4), and $(0,\xi_L)$ transforms as a $\bf \bar{6}$ of SU(6).
Similarly, for right-handed quarks:
\beq
y_R f \, (\xi_R,0)\cdot \tilde{\Psi} = y_R f\,  \left( 1 - \frac{i}{\sqrt{2} f_6} \pi_8^a \lambda^a \right) \xi_R U_4^{(R)} \psi - y_R f\, \frac{i}{\sqrt{2} f_6}  \pi_6 \xi_R U_4^{(R)}  \eta  + \dots \label{eq:rightpreY}
\eeq
Note that the colour indices are omitted, and that the interactions with the un-coloured pions, the Higgs and the singlet $\eta$, arise form the expansion of the pion matrix $U_4^{(R)}$. 
This shows that the couplings of the octet and sextet are proportional to the pre-Yukawa couplings $y_{L/R}$ in the fermion sector.

As an explicit example, we will consider a set of composite fermions transforming as $\Psi = ({\bf 6}, {\bf 6})$ of SU(4)$\times$ SU(6), which decomposes as a ${\bf 5}_{\Sp(4)}$ and a ${\bf 1}_{\Sp(4)}$:
\beq
\Psi_5 = ({\bf 5}, {\bf 6})_{\rm Sp(4)\times SO(6)}\,,\qquad \Psi_1 = ({\bf 1}, {\bf 6})_{\rm Sp(4)\times SO(6)}\,.
\eeq
This choice corresponds to mixing the tops with the composite baryon $\psi_1$, or $\psi_2$.
In the vacuum where the EW symmetry is unbroken, the two composite fermions can be written in terms of an antisymmetric matrix in the Sp(4) space as follows:
\beq
\psi_5 &=& \left( \begin{array}{cc}
\frac{1}{2} \tilde{T}_5 i \sigma_2 & \frac{1}{\sqrt{2}} Q \\
- \frac{1}{\sqrt{2}} Q^T & \frac{1}{2} \tilde{T}_5 i \sigma_2
\end{array} \right)\,, \quad Q = \left( \begin{array}{cc}
X_{5/3} & T \\
X_{2/3} & B
\end{array} \right)\,; \\
\eta_5 &=& \left( \begin{array}{cc}
\frac{1}{2} \tilde{T}_5^c i \sigma_2 & \frac{1}{\sqrt{2}} Q^c \\
- \frac{1}{\sqrt{2}} {Q^c}^T & \frac{1}{2} \tilde{T}_5^c i \sigma_2
\end{array} \right)\,, \quad Q^c = \left( \begin{array}{cc}
- X_{2/3}^c & B^c \\
X_{5/3}^c & -T^c
\end{array} \right)\,; 
\eeq
where the fields with a ${}^c$ are the change-conjugate of the right-handed chiralities. Note also that $\psi_5$ is a colour $\bf3$ with hypercharge $2/3$, while $\eta_5$ is a colour $\bf\bar{3}$ with hypercharge $-2/3$.
The singlet can be written as:
\beq
\psi_1 = \frac{1}{2} \tilde{T}_1 \left(\begin{array}{cc}
i \sigma_2 & 0 \\
0 & - i \sigma_2
\end{array} \right)\,, \quad \eta_1 = \frac{1}{2} \tilde{T}_1^c \left(\begin{array}{cc}
i \sigma_2 & 0 \\
0 & - i \sigma_2
\end{array} \right)\,.
\eeq
With this parametrisation, the masses of the two fermions can be written as
\beq
& M_5\, \mbox{Tr} [\eta_5 \cdot \Sigma_B \cdot \psi_5 \cdot \Sigma_B] + M_1\, \mbox{Tr} [\eta_1 \cdot \Sigma_B \cdot \psi_1 \cdot \Sigma_B]  + h.c.= & \\
& M_5 (B^c B + T^c T + X_{2/3}^c X_{2/3} + X_{5/3}^c X_{5/3} + \tilde{T}_5^c \tilde{T}_5) + M_1 \tilde{T}_1^c \tilde{T}_1 + h.c.&
\eeq
where
\beq
\Sigma_B = \left(\begin{array}{cc}
i \sigma_2 & 0 \\
0 & - i \sigma_2
\end{array} \right)
\eeq
is introduced to properly contract the Sp(4) indices. The elementary SM fermions can be embedded in a spurion similar to $\psi_5$ for the left-handed doublets 
$(t_L, b_L)$ and $\eta_1$ for the right-handed singlet $t_R^c$.
We can now simply plug these matrices in the above formulas and expand in the SU(4)/Sp(4) pseudo-Goldstone mesons (see Appendix~\ref{app:1} for details).
Putting the results together, in the basis $\{ t, T, X_{2/3}, \tilde{T}_1, \tilde{T}_5\}$, the mass in the top sector is given by:
\beq
M_{\rm top} = \left( \begin{array}{ccccc}
0 & y_{5L} f \cos^2 \frac{\epsilon}{2} & - y_{5L} f \sin^2 \frac{\epsilon}{2} & \frac{y_{1L} f}{\sqrt{2}} \sin \epsilon & 0 \\
\frac{y_{5R} f}{\sqrt{2}} \sin \epsilon & M_5 & 0 & 0 & 0 \\
\frac{y_{5R} f}{\sqrt{2}} \sin \epsilon & 0 & M_5 & 0 & 0 \\
y_{1R} f \cos \epsilon & 0 & 0 & M_1 & 0 \\
0 & 0 & 0 & 0 & M_5 
\end{array} \right)\,.
\eeq
One interesting feature is that the singlet $\tilde{T}_5$ does not mix with the other fields in the mass matrix. Additional couplings of the Higgs, in the form
\begin{equation}
i c_{L,R}\bar{\psi}_{5L,R}d_\mu\gamma^\mu\psi_{1L,R}=c_{L,R}\frac{\partial_\mu h}{2f}(\bar{T}_{L,R}+\bar{X}_{2/3L,R})\gamma^\mu\tilde{T}_{1L,R}\,,
\end{equation}
also do not explicitly involve $\tilde{T}_5$, thus we conclude that it does not mix with elementary fields at tree level even in the true mass basis. Therefore 
our set-up is similar to the minimal case $\SO(5)/\SO(4)$ with ${\bf 1}+{\bf 4}$ top partners (known as MCHM5~\cite{Contino:2006qr,Agashe:2006at}). Once Yukawa couplings for the light quarks are turned 
on, the analysis carried on in \cite{Cacciapaglia:2015dsa} can be repeated here without major modifications: besides minor differences in the form of order one 
coefficients, as for instance the deviations of the couplings of the physical Higgs to fermions, the mass of the top can be obtained without violating flavour bounds.

Going in the mass eigenstate basis, at leading order in $v/f$, the couplings to top and bottom of the coloured mesons are:
\beq
i g_{\pi_8 t_L t_R^c} &=& \frac{m_{\rm top}}{f_6} \frac{2+\cos (2\phi_L) + \cos (2 \phi_R)}{2 \sqrt{2}} + \dots\,, \\
i g_{\pi_6 t_R^c t_R^c} &=&  \frac{M_1}{f_6} \frac{\sin^2 \phi_R}{\sqrt{2}} + \dots\,, \\
i g_{\pi_6^c t_L t_L} &=& 0 + \dots\,, \\
i g_{\pi_8 b_L b_R^c} &=& 0\,;
\eeq
where
\beq
\tan \phi_L = \frac{y_{5L} f}{M_5}\,, \qquad \tan \phi_R = \frac{y_{1R} f}{M_1}\,.
\eeq
The above coupling respect the form we anticipated in \refeq{eq:tcoup}.

As the pre-Yukawa couplings are required to be $\mathcal{O} (1)$ in order to obtain a large enough top mass, from the above results we can see that the sextet has $\mathcal{O} (1)$ couplings to the right-handed tops (mass eigenstates), while the couplings to the left-handed tops are suppressed by $v^2/f^2$.
On the other hand, the couplings of the octet are suppressed by the top mass over the condensation scale.
This result can be easily understood in terms of the EW quantum numbers of the states: the only gauge invariant combination of fermions that can form a sextet with charge $4/3$ is a bilinear in the right handed tops, while all the other couplings need to be generated via the electroweak symmetry breaking source.

\subsection{Phenomenological considerations}

As outlined earlier in this section, the model under consideration contains a large number of composite resonances: scalars, fermions and spin-1. Such states have been widely considered in phenomenological studies, however the interplay between them, in particular the scalar mesons and the fermionic baryons which are expected to be lighter, is still a fairly unexplored land.

The model we consider here, for instance, contains two un-coloured pNGBS: $\eta$ and the singlet associated to the broken U(1) flavour symmetry. 
 The phenomenology of the $\eta$ has been discussed in Ref.\cite{Arbey:2015exa} in the absence of top partners. However, a natural expectation is that it is lighter than the top partners, thus they can decay into a SM quark plus $\eta$, providing final states not yet considered in experimental studies. A recent attempt to investigate the impact of $\eta$ on the top partner searches has been presented in~\cite{Serra:2015xfa}.
 Similar impact may be expected from the singlet.
 
The coloured pNGBs are expected to have a similar mass as the top partners, thus their phenomenology crucially depends on the mass hierarchy. 
When they are lighter than the top partners, they may open a new decay mode for the top partners, thus affecting their phenomenology. For an inverted hierarchy, their decay  into top partners become kinematically allowed, thus providing additional channels for single production of top partners.
In the rest of this paper we will focus on the case where the mesons are lighter than the top partners, and we will only consider their direct production and decays into SM quarks.
 In Section \ref{sec:bound} we develop an effective field theory description which captures the main interactions of the sextet and octet states arising for example in the composite  model discussed here. In Sec.~\ref{sec:LHC} we study the LHC phenomenology of the model.
 We then leave the investigation of their interplay with top partners to future work.
 
 It should be noted that not all baryons need to couple to the SM quarks, so the possibility that some of them couple directly to a dark sector, not included in the model we present here, is viable. This case has been studied in general in~ \cite{Anandakrishnan:2015yfa}.

\section{Sextet effective theory}
\label{sec:bound}

Studying the bound states of the FCD model with top partners in the previous section, we found that, apart from the baryonic top partners, other states with potentially interesting collider signatures
can be present in the spectrum. Remarkably, one has scalar coloured mesons whose mass is expected at the same, if not lower, scale as the coloured baryons.
In the specific model, the spectrum contains a complex colour sextet (with charge $Q=4/3$) and a real colour octet.
The presence of such states is rather generic, as any dynamics that generates coloured baryonic bound states is expected  to contain also coloured mesons. In the rest of this paper, we will mainly focus on the possibility that such states are lighter than the fermionic states, thus they can only decay directly to a pair of SM particles.

In order to keep the discussion generic, we will study the constraints on a general Lagrangian.
The couplings of the sextet and octet can be guessed by looking at the invariance under colour and charge: a sextet can only be obtained by combining two quarks, as for SU(3) representations ${\bf 3} \otimes {\bf 3} \supset {\bf 6}$, or four anti-quarks ${\bf \bar{3}} \otimes {\bf \bar{3}} \otimes {\bf \bar{3}} \otimes {\bf \bar{3}} \supset {\bf 6}$.
Conversely, the octet can only couple to a quark-antiquark pair .

\subsection{Baryon number conserving Lagrangian and couplings}

An effective Lagrangian for the sextet and octet scalars can be easily build, by imposing invariance under colour and electric charge: we focus here on the case $x=2/3$, for which the sextet has charge $4/3$ and thus couples to a pair of up-type quarks. Below the EWSB scale, the effective Lagrangian reads
\beq
\mathcal{L} & = &{|D_\mu\pi_6 |}^2 - m_{\pi_6}^2 {|\pi_6|}^2 + \frac{1}{2} {(D_\mu \pi_8)}^2 - \frac{1}{2} m_{\pi_8}^2 {(\pi_8)}^2 - V_{\rm scalar} (\pi_6, \pi_8) \nonumber \\
& & + a_R ~ \pi_6 t_{R}^c t_{R}^c + a_L ~ \pi_6^c t_{L} t_{L} + b~ \pi_8 t_R^c t_L + h.c. \label{eq: lag sextet}
\eeq
where $V_{\rm scalar}$ contains generic self-interactions between the scalars, and $t_{L/R}$ are chiral Weyl spinors (${^c}$ indicates the charge conjugation).
Parity is in general not conserved in \refeq{eq: lag sextet}, because only the coupling $a_R$ corresponds to a gauge invariant operators, while the other couplings can only be generated via the EW symmetry breaking. In fact, one can expect
\beq
\frac{a_L}{a_R} \sim \mathcal{O} (v^2/\Lambda^2)\,, \quad \frac{b}{a_R} \sim \mathcal{O} (v/\Lambda)\,,
\eeq
where $\Lambda$ is the scale of new physics. This hierarchy of couplings is reflected in the explicit model discussed in the last section, where there $\Lambda = f$.

This Lagrangian is not the most general one because we couple the sextet only to the third generation quarks, while in principle we should take $a_{L,R}$ and $b$ to be $3\times3$ matrices in family space:  this choice can be motivated from a UV point of view, if top partners, and partial compositeness, is only invoked to give mass to top and bottom, while the light quarks, and leptons, receive their mass via bi-linear couplings.
The robustness of this setup with respect to the experimental bounds on flavour observables has been studied in \cite{Cacciapaglia:2015dsa}, where it was shown that this scenario passes all the bounds, without requiring special symmetries, independently on the coset and representations of the composite fermions:
under the assumption of large enough anomalous dimensions of the bi-linear operators a high flavor scale is compatible with charm and bottom masses of the correct size, assuming also a near conformal behaviour between the two scales.
The couplings to the coloured mesons are only generated for the top (and bottom) via the same mechanism giving them mass, while a coupling to the light quarks is induced once the quark fields are rotated in the true mass eigenbasis.
However, a natural hierarchy between the contribution of the bi-linear couplings and the top mass ensues.
The couplings in the mass basis, therefore, will be expressed in terms of mixing matrices of the form
\beq
a_{L}\, V_{uL}^{i3} V_{uL}^{j3}\,,\quad a_{R}\, V_{uR}^{*i3} V_{uR}^{*j3}\,,\quad b\, V_{uR}^{*i3} V_{uL}^{j3}\,,\quad\mbox{with}\quad
i,j = 1,2,3\,.
\eeq
The rotation matrices $V_{uL,R}$ are intrinsically hierarchical, due to the fact that the charm mass is generated by bi-linear interactions, while the top mass comes from partial compositeness, thus we can estimate \cite{Cacciapaglia:2015dsa}
\beq\label{eq:V hier}
V_{uL,R}=\left(\begin{array}{ccc}
O(1)&O(1)&O(\frac{m_c}{m_t})\\
O(1)&O(1)&O(\frac{m_c}{m_t})\\
O(\frac{m_c}{m_t})&O(\frac{m_c}{m_t})&O(1)
\end{array}\right)\,.
\eeq
The couplings of the coloured mesons to light quarks are therefore strongly suppressed by powers of $m_c/m_t$, and are thus irrelevant for the phenomenology of these states at the LHC.

\subsection{Flavour bounds}
\label{sec:fla}

The exchange of coloured mesons can in principle generate large flavour changing neutral currents in the up-sector: the most dangerous one being the sextet which has couplings to the right-handed tops which is not suppressed by $v/f$. Flavour aspects in presence of scalar diquarks have been extensively investigated in \cite{Giudice:2011ak}
.
The tree level exchange of $\pi_6$ scalars leads to four quark interactions. The induced operators has the form
\begin{equation}
\frac{{|a_R|}^2}{m_{\pi_6}^2}(t_R^c t_R^c)(t_R t_R)=\frac{{|a_R|}^2}{2m_{\pi_6}^2}(t_R^c\sigma^\mu t_R)(t_R^c\sigma_\mu t_R)\,,
\end{equation}
where we used a Fierz-identity, and the colour indices are left understood. After rotating in the mass eigenstates, the operator $(c_R^c\sigma^\mu u_R)(c_R^c\sigma_\mu u_R)$, violating flavour number by two units, has a non zero coefficient given by
\begin{equation}
\frac{{|a_R|}^2}{m_{\pi_6}^2}{\left({V_{uR}^{23}}^*\right)}^2{V_{uR}^{13}}^2\sim\frac{10^{-9}}{{(1\mbox{ TeV})}^2}{\left(\frac{{|a_R|}^2}{1}\right)}{\left(\frac{1\mbox{ TeV}}{m_{\pi_6}}\right)}^2\,.
\end{equation}
The coefficient is experimentally bound to be less than $10^{-7}$ at the fixed reference scale of $1$ TeV  \cite{Calibbi:2012at}, thus giving a mild $a_R$-dependent bound on the mass of the sextet
\beq
m_{\pi_6} > 0.1\, |a_R|\, \mbox{TeV}\,.
\eeq
 An analogous operator is found with $R\rightarrow L$ and the experimental bound for it is of the same order \cite{Calibbi:2012at,Isidori:2010kg}.

A further constraint on the flavour-conserving part of the operator can be obtained from the study of the angular distribution of dijet events at the LHC, which  constrains the size of a generic four quark interaction. For operators 
\begin{equation}
\frac{1}{M^2}(\bar{q}_X\gamma^\mu q_X)(\bar{q}_X\gamma_\mu q_X)\,,\quad q=u,d\,,\quad X=L,R\,,
\end{equation}
$M>O(1)$ TeV is required \cite{Bazzocchi:2011in,Domenech:2012ai}, where the precise value depends on the details of the operator. 
The operator generated in the model discussed in the last section is well below the bound thanks to the suppression from flavour mixing.

\subsection{Baryon number violation and neutron-antineutron oscillations}

The model we introduced in Section~\ref{sec:model} is baryon number conserving: is suffices in fact to assign baryon number $3 B = 1$ ($-1$) to the fundamental fermions $\chi$ transforming as a colour triplet (antitriplet), which translates in the sextet having baryon number $2/3$ and the octet carrying no baryon number.
In general, however, one can write an additional operator coupling the sextet to down-type quarks:
\beq
\Delta \mathcal{L}_{\rm eff} = c_R~ \pi_6 b_R b_R b_R b_R   + h.c. \label{eq:BNV}
\eeq
which is also unsuppressed by powers of $v/\Lambda$. The above operator has total baryon number $2$ (note also that $c_R$ has mass dimension $-3$). It cannot thus mediate the decay of the proton, however it will induce neutron-antineutron oscillations.

The current experimental limit, set by Super Kamiokande \cite{Abe:2011ky}, on the period of neutron antineutron oscillations in empty space\footnote{This measure is extracted looking at oscillations in nuclei. The conversion to the free case is obtained estimating a nuclear suppression factor.} is
\begin{equation}\label{eq: Super K}
\tau_{n-\bar{n}}\geq2.44\cdot10^8\,\, s\,\,\,\mbox{at $90\%$ C.L.}
\end{equation}
This translates into a bound $\delta m\leq10^{-33}$ GeV on the off-diagonal $n\bar{n}$ term in the effective $2\times2$ Hamiltonian for neutron oscillations \cite{Mohapatra:2009wp}. In the SM Lagrangian the dimension nine operator
\begin{equation}\label{eq: n nbar nine}
\frac{c_{\Delta B=2}}{M^5}uddudd
\end{equation}
translates to 
\begin{equation}
\delta m\sim c_{\Delta B=2}\Lambda_{QCD}{\left(\frac{\Lambda_{QCD}}{M}\right)}^5=O(10^{-20}\mbox{ GeV})\left(\frac{c_{\Delta B=2}}{1}\right){\left(\frac{1\mbox{ TeV}}{M}\right)}^5\,.
\end{equation}

In the case under study, the coupling in \refeq{eq:BNV}, together with the Lagrangian \refeq{eq: lag sextet}, violates baryon number: integrating out at tree level the complex sextet $\pi_6$, and rotating in the mass eigenstate basis, we get, among others, the following dimension nine operator\footnote{Here, $\alpha,\beta=1,2$ are spinor indices 
and $i,j=1,2,3$ ${\SU(3)}_c$ indices.}
\begin{equation}
-\frac{a_R^*c_R^*}{m_{\pi_6}^2}{(u_R)}_{\dot{\zeta}}^p{(d_R)}_{\dot\alpha}^i{(d_R)}_{\dot{\gamma}}^k{(u_R)}_{\dot\eta}^q{(d_R)}_{\dot\beta}^j{(d_R)}_{\dot\delta}^l\epsilon^{\dot\alpha\dot\beta}\epsilon^{\dot\gamma\dot\delta}\epsilon^{\dot\zeta\dot\eta}\epsilon_{ikp}\epsilon_{jlq}{(V_{uR}^{13})}^2{(V_{dR}^{13})}^4\,,
\end{equation}
which is of the form of \refeq{eq: n nbar nine} with
\begin{equation}
\frac{c_{\Delta B=2}}{M^5}=-\frac{a_R^*c_R^*}{m_{\pi_6}^2}{(V_{uR}^{13})}^2{(V_{dR}^{13})}^4\,.
\end{equation}
The hierarchic structure in the mixing matrices, \refeq{eq:V hier}, implies then
\begin{equation}
\delta m=O(10^{-33}\mbox{ GeV}){\left(\frac{a_R}{1}\right)}{\left(\frac{c_R}{{(1\mbox{ TeV})}^{-3}}\right)}{\left(\frac{1\mbox{ TeV}}{m_{\pi_6}}\right)}^2\,,
\end{equation}
compatible with the bound \refeq{eq: Super K} for $m_{\pi_6}\sim1$ TeV and order one couplings.

\section{LHC phenomenology of the  sextet and octet}
\label{sec:LHC}

\begin{figure}[tb!]
\includegraphics[width=1.0\textwidth]{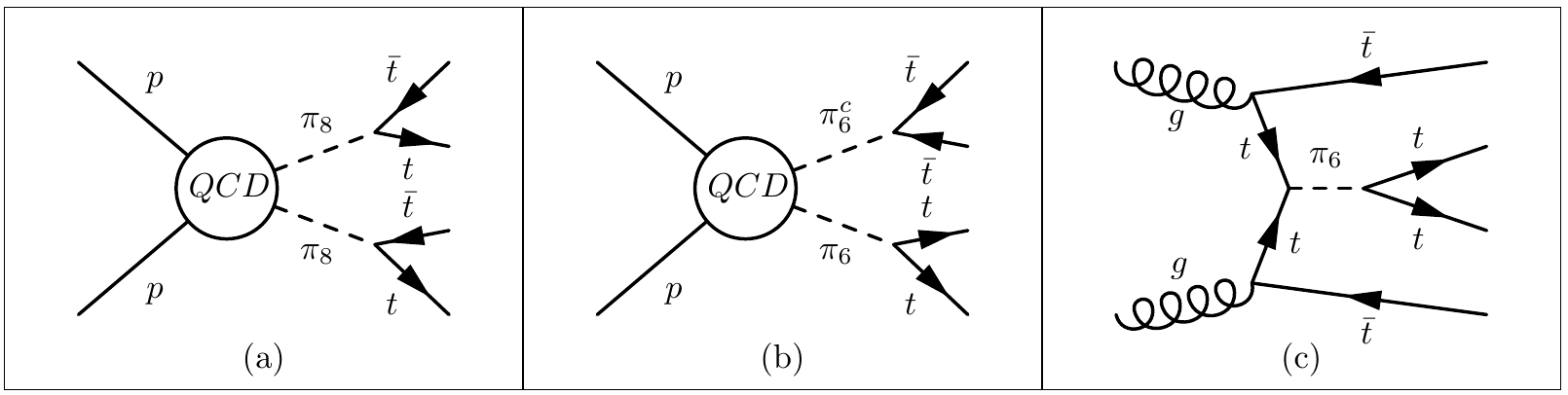}
\caption{$tt\bar{t}\bar{t}$ production channels involving sextets and octets. The diagrams (a) and (b) show QCD pair production channels, where the QCD vertex includes the $gg\pi^{\,c}_{8,6}\pi_{8,6}$ interaction as well as an $s$-channel gluon (with $gg$ or $q\bar{q}$ initial state) or a  $t$-channel $\pi_{8/6}$ exchange with a $gg$ initial state. Fig. (c) shows the $tt\bar{t}\bar{t}$ contribution from $\pi_6$ single production. The single $\pi_6^c$ production contribution is obtained analogously (with  $t\leftrightarrow \bar{t}$ and $\pi_6\rightarrow \pi_6^c$).}\label{fig:channels}
\end{figure}

The Run II at the LHC is an extremely important opportunity to test the presence of exotic particles predicted by the class of composite models we are considering, especially regarding coloured states. Early stage studies for pair and single production of the coloured scalars (triplet, sextet and octet) can be found in the literature~\cite{Chen:2008hh,Han:2009ya, Berger:2010fy}. Here we will focus on the case where the coloured states couple mainly to tops, which is relatively easy to probe at the LHC. Four top quark events are a particularly attractive channel, as the production rates in the SM are very small (of the order of 1 fb at $ \sqrt s = 8$ TeV for the LHC), while they can be significantly enhanced in extensions of the SM, as studied for various particle physics scenarios ~\cite{Battaglia:2010xq,Gregoire:2011ka,Cacciapaglia:2011kz,AguilarSaavedra:2011ck,Deandrea:2014raa}
and searched for by ATLAS and CMS ~\cite{ATLAS:2012hpa,CMS:2013xma, ATLAS:2013jha, Khachatryan:2014sca, Aad:2015gdg, Aad:2015kqa}.

\subsection*{Existing bounds from LHC Run I}

\begin{figure}[tb!]
\center
\includegraphics[scale=0.63]{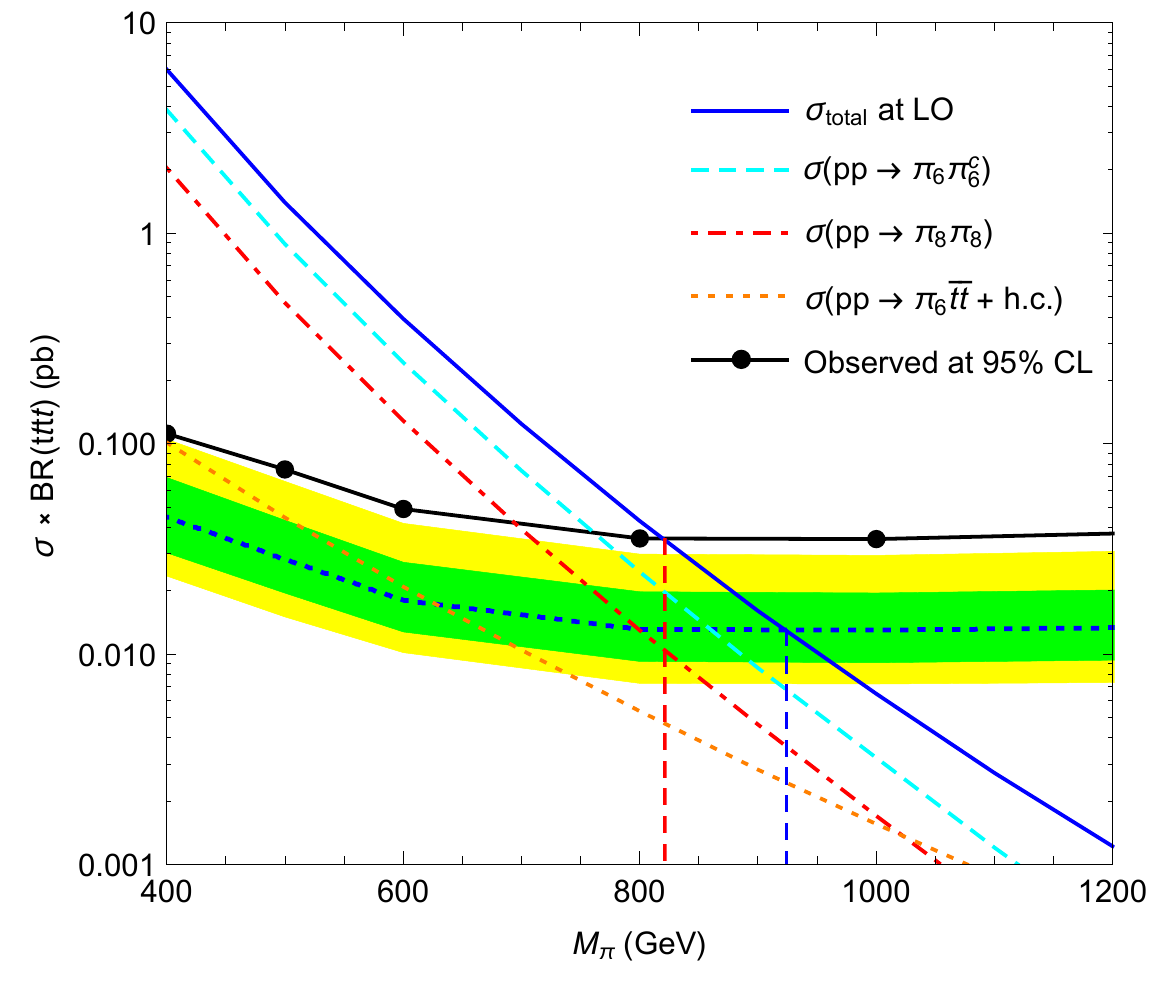}
\includegraphics[scale=0.64]{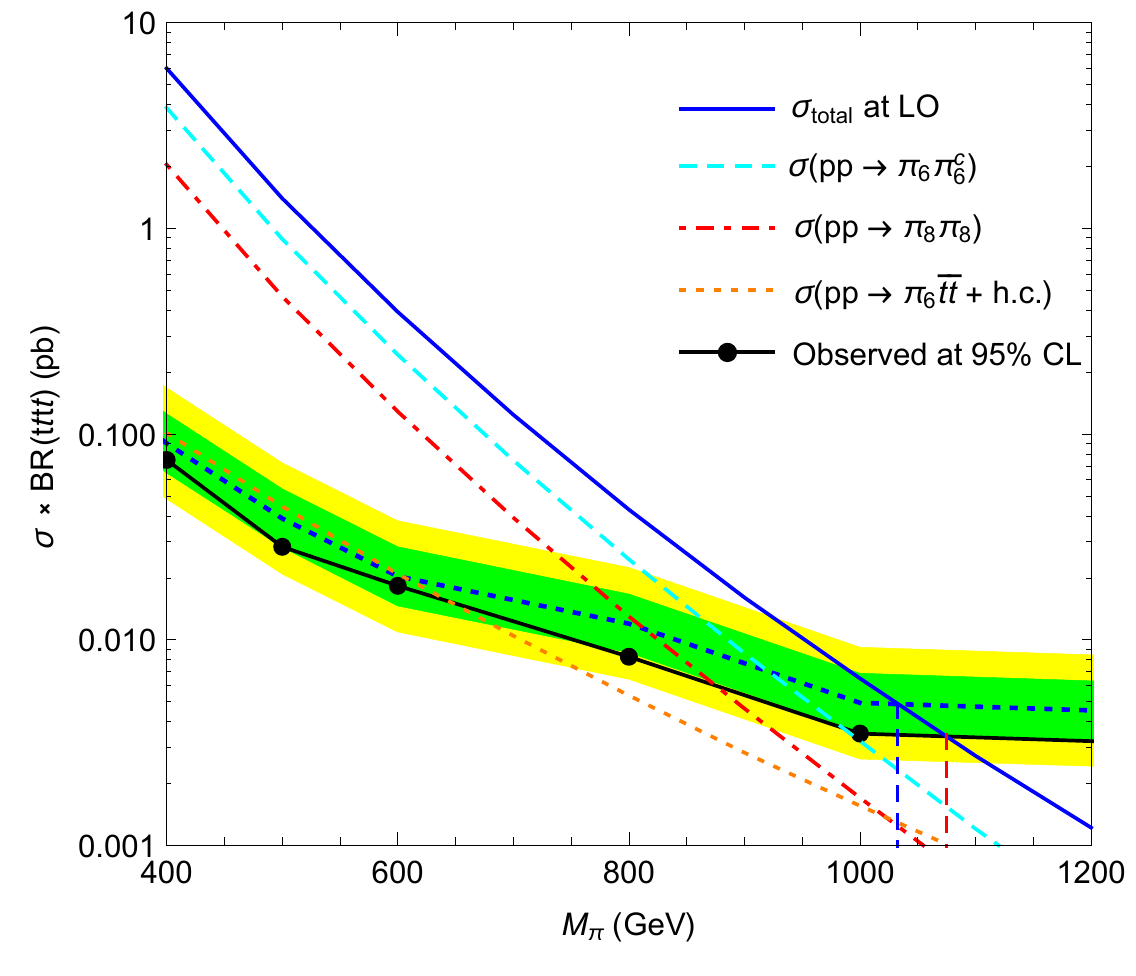}
\caption{Cross sections for the sextet and octet scalars at the LHC at $8$ TeV, with $a_R =1$. Left panel: comparison with the ATLAS 2SSL search~\cite{Aad:2015gdg}, where the green (yellow) band is for $1 \sigma$ ($2 \sigma$) expected limit and the solid black curve is the observed limit. Right panel: comparison with the ATLAS 1-lepton search observed limit~\cite{Aad:2015kqa}. }
\label{cs}
\end{figure}

Given the  specific couplings predicted in the composite model we consider,  there are four main channels contributing to the 
$tt \bar t \bar t$ final state, which include the single and pair sextet production, i.e. $ p p \to \bar t \bar t \pi_6 $, $t t \pi_6^c$, $\pi_6 \pi_6^c $, with $\pi_6 \to t t$,  and also the pair octet production, i.e.  $p p \to \pi_8 \pi_8 $, with $\pi_8 \to t \bar t $ ({\it c.f.} Fig. \ref{fig:channels}).\footnote{We neglect here the single octet production
$ p p \to t \bar t \pi_8$, as this process is suppressed by a factor of  $m_t^2/ M_1^2 \sim \mathcal{O}(10^{-2})$ from the octet coupling, compared with the process with single sextet production.} We first use the analysis of same sign dilepton (2SSL) at 8 TeV to extract a bound on the mass of the coloured scalar resonances, assuming equal masses $m_{\pi_6}\approx m_{\pi_8}\approx M_\pi$. The cross section at the leading order for each channel at a 8 TeV LHC is shown in the left panel of  Figure~\ref{cs}, together with the experimental bound. Considering the contribution from all the important channels, a sextet and octet with masses smaller than $\sim 800$ GeV are disfavoured by the 2SSL analysis, done with the full Run I data of $20.3 ~\mbox{fb}^{-1}$, by the ATLAS collaboration~\cite{Aad:2015gdg}. The recent ATLAS search~\cite{Aad:2015kqa} in the lepton-plus-jets final state, also on the full Run I data, yields a more stringent bound for the 4$t$ cross section: comparing this bound with the model cross sections, as shown in the right panel of Fig. \ref{cs}, implies a limit of  $M_{\pi} \gtrsim 1.1$ TeV.
The single-lepton analysis, however, relies on a shape fit on a kinematical distribution ($H_T$) in a region where very few events are present. We therefore decided to rely on the 2SSL analysis, which is a robust cut-and-count search, to set the lowest acceptable mass value for the coloured scalars.\footnote{CMS performed a search for four tops in the lepton-plus-jets final state which yields an upper bound of $32~\mbox{fb}^{-1}$ at 95\% c.l. on the four-top production cross section \cite{Khachatryan:2014sca}. As this study focusses on SM-like  four-top signatures, the bounds from it on colour sextets and octets are weaker than the one of the ATLAS study used here. CMS also performed a search in the 2SSL and $b$-jets channel \cite{Chatrchyan:2012paa}
 and the 2SSL and jets channel \cite{Chatrchyan:2013fea}, which are tailored for supersymmetric model signatures and yield upper bound of  $49~\mbox{fb}^{-1}$ at 95\% c.l. on the four-top production cross section. This bound resembles the ATLAS search bound of \cite{Aad:2015gdg}, although the bounds cannot be compared directly as they have been determined based on different underlying model assumptions.}
In the comparisons of the cross section in Fig.~\ref{cs} to the ATLAS studies, we are assuming that the efficiency of the signal selection is the same as the one obtained by ATLAS on the octet pair production signal: in the 2SSL analysis, one might expect some differences as the leptons arising from $\pi_6 \rightarrow t t$ decays have different kinematics. For the single lepton analysis the assumption of same efficiencies is fully justified as the search is blind to the charge of the $t$ vs. $\bar{t}$ such that $\pi_8\rightarrow t \bar{t}$ and $\pi_6\rightarrow t t $ do not lead to distinguishable topologies for this search.

\begin{figure}[tb!]
\center
\includegraphics[scale=0.65]{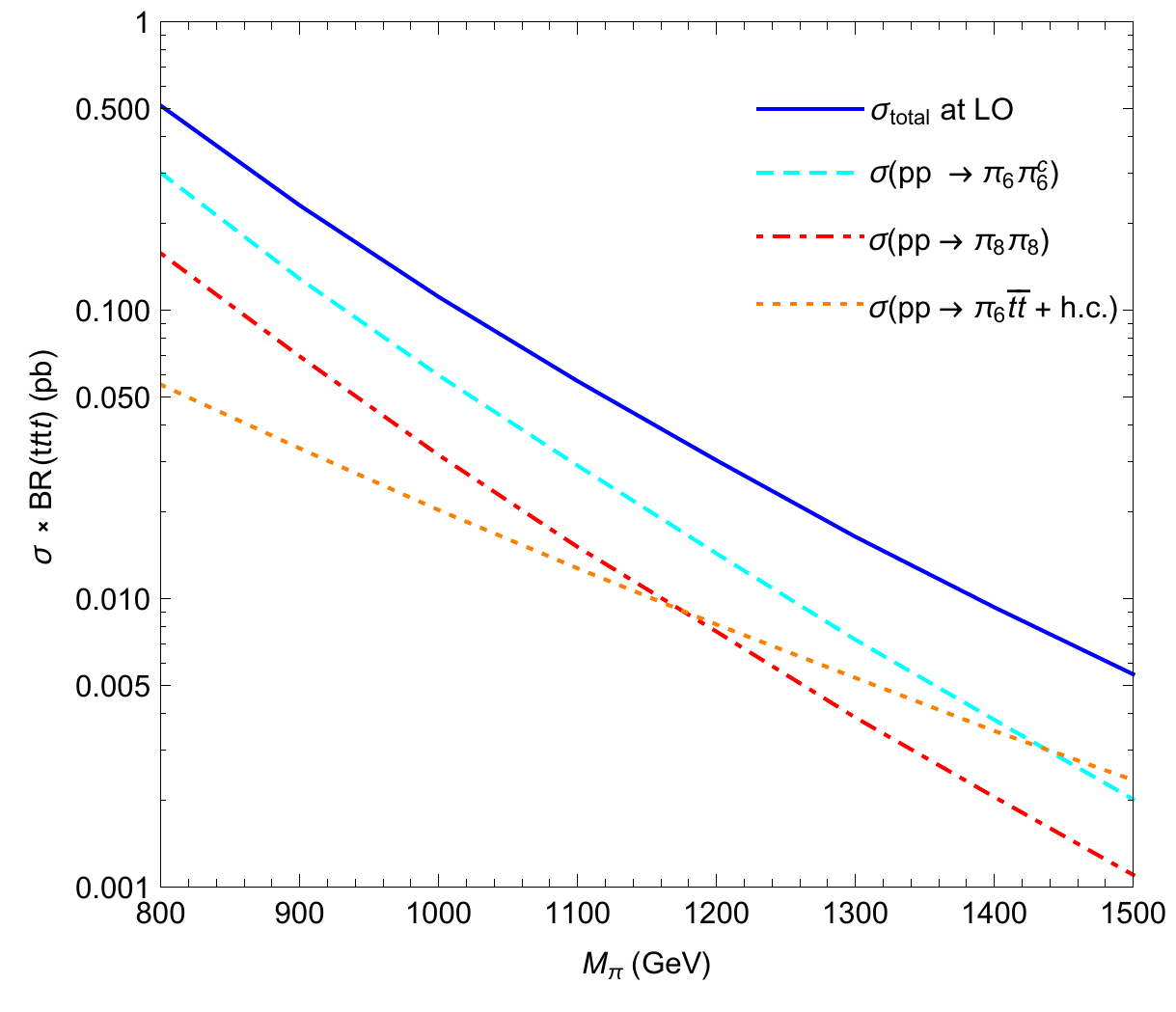}
\caption{Cross sections for the sextet and octet scalar production at the LHC $13$ TeV, with $a_R =1$. }
\label{cs13}
\end{figure}

The leading order cross sections at the Run II LHC  with $\sqrt s = 13 $ TeV is shown in Figure~\ref{cs13}, with the $\pi_6$-$t$-$t$ coupling set to one. As can be seen, for this coupling value the single $\pi_6$ production cross section becomes larger than the octet $\pi_8$ pair production cross section at  $M_\pi > 1.2$ TeV.  Note that in computing the cross sections of  Figure~\ref{cs} and~\ref{cs13}, and all the numerical results in this section, we are using results at leading order. Next-to-leading order corrections, expressed in terms of k-factors, for the octet pair production have been calculated in \cite{Degrande:2014sta}, and are found to be close to unity for $M_\pi$ above a TeV. The k-factors for the sextet single and pair production are not available in the literature. For pair production of the sextet, the k-factors can be expected to be similar to the octet ones, i.e. close to unity, such that our pair production channels should be well approximated when treating them at leading order. For the $\pi_6$ single production, k-factors may be very different: however, this production channel depends on the free coupling $\pi_6$-$t$-$t$ such that -- at least for the overall cross section -- a k-factor can be effectively absorbed into the definition of the coupling $a_R$ appearing in the leading order calculation.

\subsection*{Simulation for LHC Run II}

Both the single and pair production of $\pi_6$ and $\pi_8$ yield $tt\bar{t}\bar{t}$ final states for which -- as it has been found at the LHC Run I -- the 2SSL and the single lepton-plus-jets channels provide strong discovery potential. To distinguish sextet and octet signals, however, the 2SSL channel is more promising. The sextet decays to a same sign $tt$ pair, while the octet decays to $t\bar{t}$. Thus, in a 2SSL search, the two same sign leptons arise either from the decay of the same particle ($\pi_6$ or $\pi_6^c$) or from two different particles (in the case of $\pi_8\pi_8\rightarrow t\bar{t}t\bar{t}$), implying very different kinematical distributions for the 2SSL in the two cases. In the following we outline a cut-scheme to isolate sextet and octet searches from SM background and then explore how kinematic differences of sextet and octet signatures might be used in order to disentangle $\pi_6$ and $\pi_8$, in case an excess is observed at the Run II.

\begin{figure}[tb!]
\center
\includegraphics[scale=0.37]{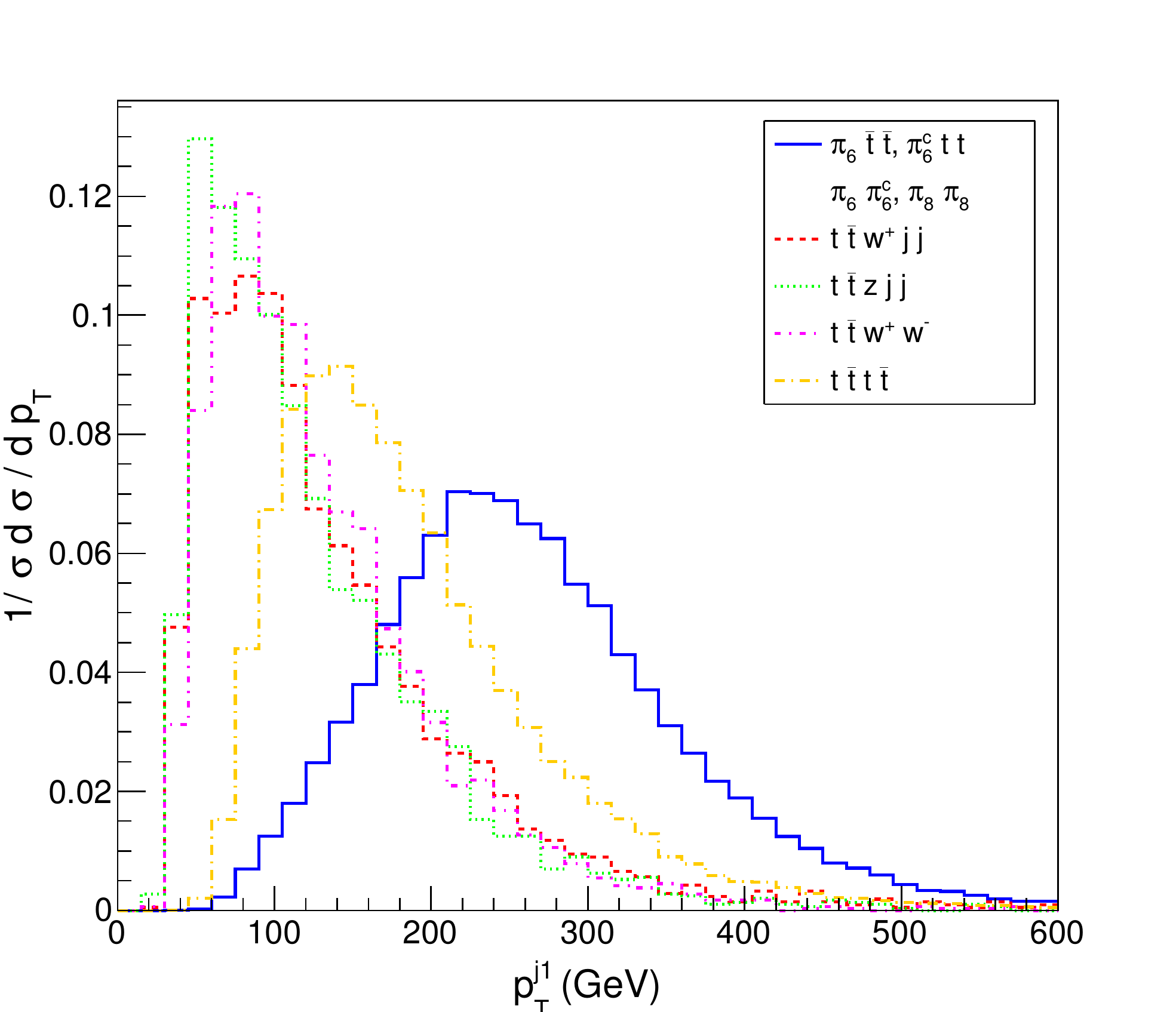}
\includegraphics[scale=0.37]{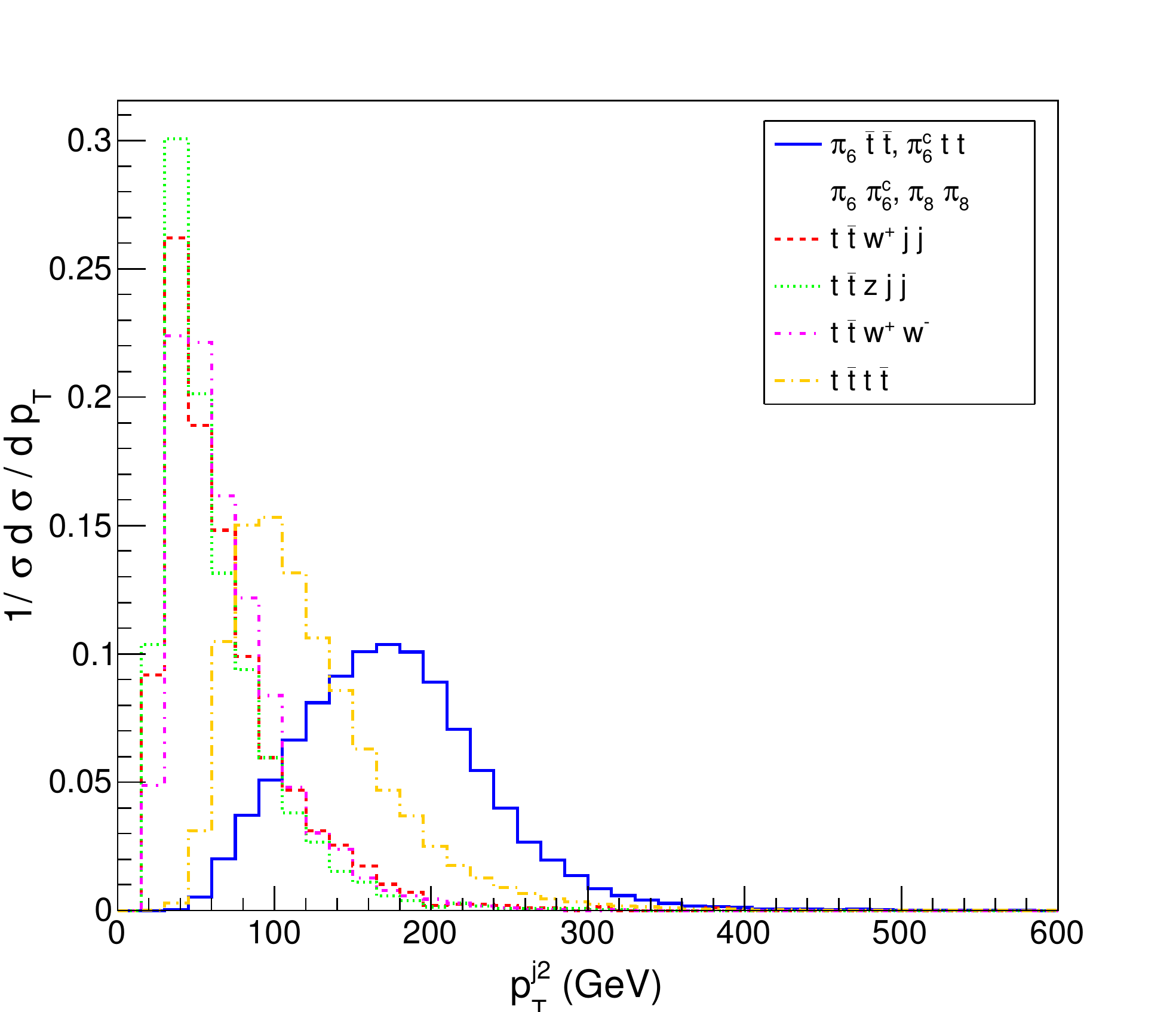}
\caption{$p_T$  distributions for the main backgrounds and for the signal (the sum of the $\pi_6 \bar{t} \bar{t}, \pi_6^ctt, \pi_6^c\pi_6,$ and $\pi_8\pi_8$ channels).  The left panel  shows the leading jet and right panel shows the subleading jet $p_T$ distribution after the basic cuts.}
\label{pt_plot}
\end{figure}

\begin{figure}[tb!]
\center
\includegraphics[scale=0.37]{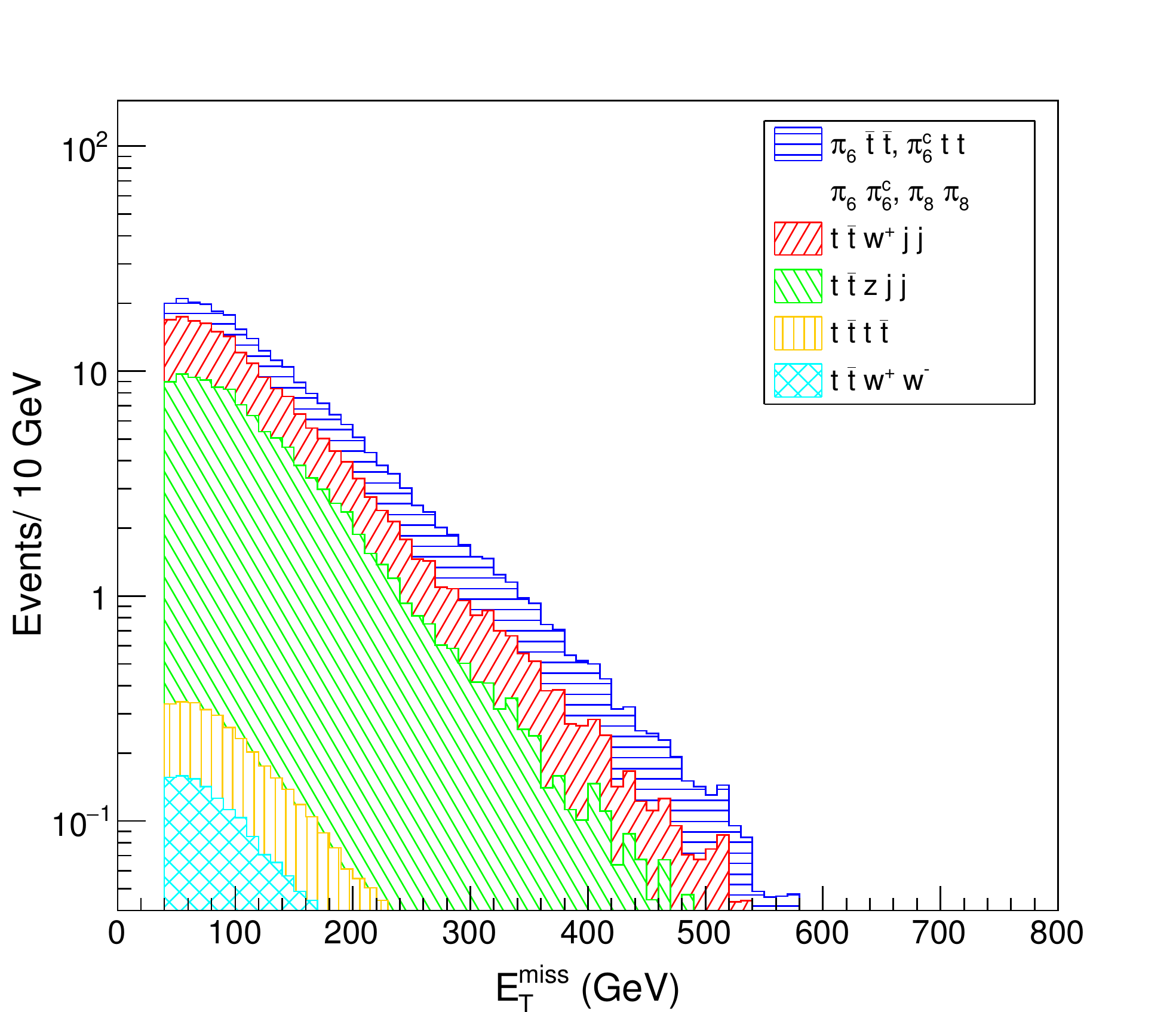}
\includegraphics[scale=0.37]{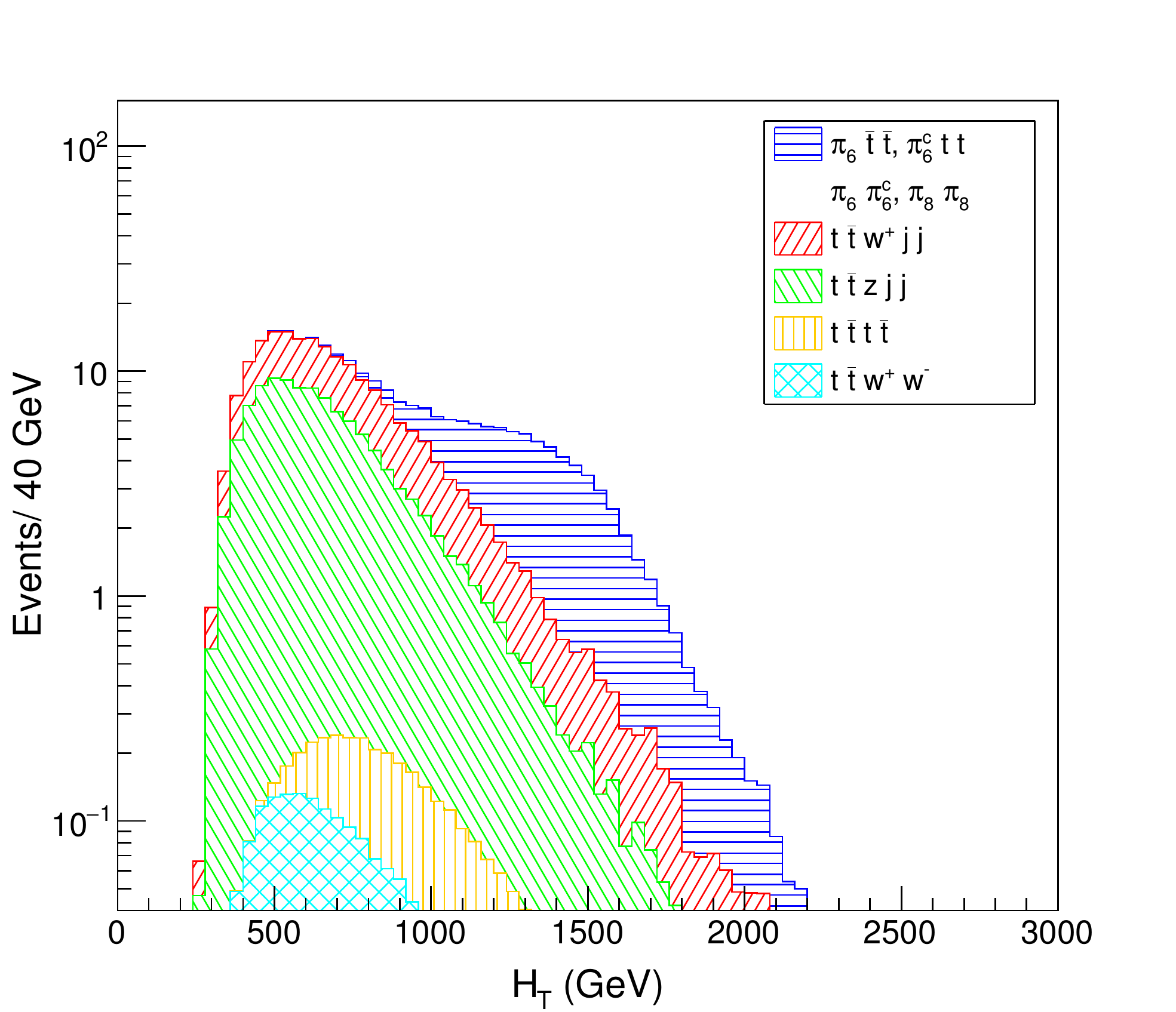}
\caption{ Left panel: staggered plot of the missing energy ($\met$)  distribution of the main backgrounds and the sum of the signal channels.\newline
Right panel: staggered plot of the $H_T$ distribution of main backgrounds and the signal.\newline
The masses of sextet and octet are assumed to be equal $M_\pi = 900$ GeV, and the coupling $\pi$-$t$-$t$ is set to be $a_R =1$. The events are selected after the basic cut, with the $b$-tag efficiency included.}
\label{Htstack}
\end{figure}

The 2SSL signature we base our study on arises from a final state with $4~ b + \ell^\pm \ell^\pm + 
4~jets +{\met}$, where the two leptons come from the leptonic decays of two same sign tops, while the remaining two tops decay hadronically. For the object selection at 13 TeV, we adopt kinematic cuts similar to the ones described in  the ATLAS search for 2SSL signature at 8 TeV ~\cite{ATLAS:2013jha},  but for simplicity we impose identical pseudo-rapidity ($\eta$) cuts on electrons and muons, without excluding the ``crack'' region $1.37 < |\eta| < 1.52$ for electrons. This simplified treatment will lead to a discrepancy in the acceptance of up to $3 \%$, which is negligible to other intrinsic uncertainties in our simulation. Moreover, we propose to include several additional critera,  e.g.  ordering the hard jets and requiring  veto cuts for an additional lepton,  in order to  increase the yield of  signal/ background and to optimise the possibility to search for  sextet and octet  scalars. The following  event-selection criteria are imposed for leptons and jets in order to perform a detailed analysis of the LHC prospect  for searching  this specific signature:
\begin{enumerate}
\item We demand at least two b-tagged jets, where we assume $ \simeq 70 \%$ tag rate for a $b$-jet, $\simeq 8 \%$ mistag rate for a $c$-jet, and a flat  $\simeq 1 \%$ mistag rate for a light quark or gluon without accounting for its $p_T$ dependence. For the signals with four top quarks (and therefore four $b$-jets), the signal efficiency of this cut is high ($\sim 92\%$) while backgrounds with two top quarks are reduced by $~50 \%$.
\item We demand 2SSL with positive charge, $l^+ l^+$, and transverse momentum $p_T^\ell > 24~\mbox{GeV}$, and pseudo-rapidity  $ \left| \eta_\ell \right|< 2.5$. We specify the 2SSL charge here because $l^+ l^+$ can only arise from two tops (but not anti-tops) which can for example arise from the decay of a $\pi_6$ (but not a $\pi_6^c$). The following discussion can be repeated for the $l^- l^-$ channel in complete analogy, when replacing tops with anti-tops, $\pi_6$ with $\pi_6^c$, $W^+$ with $W^-$, etc. 
\item We require  at least 4 additional  jets with $ p_T^j > 24~{\rm GeV}$, and  $ \left| {{\eta _j}} \right| < 2.5$.  With the jet  number cut condition   $N_j \ge 4$, we can safely ignore SM background from diboson  processes, i.e. $WZ$, $ZZ$ and $W^+ W^-$+2 jets. 
\item As separation criteria, we demand $\Delta {R_{jj}} > 0.4$, $\Delta {R_{j \ell}} > 0.4$, and $\Delta {R_{\ell \ell}} >0.4$, with $\Delta R=\sqrt{(\Delta \eta)^2 + (\Delta \phi)^2}$ being  the angular  separation between two observable particles.
\item To account for the neutrinos, we  impose a missing transverse energy cut: $\met > 40$ GeV.
\item  We order the additional jets (except the two tagged bottom quarks) in $p^j_T$ and require that the leading jet satisfies $\mbox{max}( p_T^j) > 100$ GeV, and the subleading jet satisfies $\mbox{max}( p_T^j )> 50$ GeV. In Fig.~\ref{pt_plot}, we show that this cut has a minor effect on the signal while reducing the background by $50 \%$.
\item  Finally, we demand $H_T > 650$ GeV, where  $H_T =\Sigma p_T^{\ell} + \Sigma p_T^j $ is the scalar sum of all jet and lepton transverse momenta.  The distribution of $\met$ and  $H_T$ are shown in Fig.~\ref{Htstack} which show that only $H_T$ has a good discrimination power for signal events from the SM background.
\end{enumerate}

The signal events for sextet and octet as well as the dominant SM backgrounds ($t\bar{t}W^\pm+$jets, $t\bar{t}Z+$jets, $t\bar{t}W^+W^-$, and $t\bar{t}t\bar{t}$) at 13 TeV are generated by  MadGraph 5.2 \cite{Alwall:2014hca} with the parton distribution function (PDF) MSTW2008NLO \cite{Martin:2009iq}.  The renormalisation and factorisation scales are set to $\mu_F = \mu_R = 1/2 ~ \sum_f m_f $, where we sum over the mass of the final state particles. Note that we do not consider backgrounds arising from ``fake'' leptons and charge mis-identification, as such background can only be reliably estimated from the data. We do not use our background simulation in order to estimate discovery or exclusion potentials, here, and provide the SM backgrounds only in order to motivate the cuts chosen and to determine realistic signal efficiencies. 

The numbers of events after each cut are shown in Table~\ref{cuttable}.  The row ``no cut"  corresponds to a mild kinematic cut $p_T^j > 10$ GeV for the light jets, but zero $p_T$ cuts for  both the leptons  and  bottom quarks.  The basic cuts in the second  row  include the event selection (1--5). The acceptances after passing all the selection criteria is reported in the bottom row of the table. As  can be seen, the signal acceptance decreases with increasing sextet and octet mass. In Table~\ref{efficiency} we provide a more detailed overview of the expected number of events from different signal channels for various masses and two values of the coupling $a_R$.  We checked that for $a_R \leq 1$, the narrow width approximation holds, so that smaller values of $a_R$ can be easily obtained by rescaling the yield in single production by a factor $a_R^2$. For larger values of $a_R$, the narrow width approximation starts breaking down, and one can see effects in the efficiency due to the different kinematics of the decay products.
This is illustrated at  the point  $a_R =2$: the large width affects the  kinematic  distributions so that the signal yield in the pair production is reduced, while for the single production we see that the yield is larger than the naive factor of $4$ for light masses ($M_\pi \leq 1.2$ TeV), while it is reduced at large masses.

\begin{footnotesize}
\setlength{\tabcolsep}{4pt}
\begin{table}[t!]
\begin{center}
\begin{tabular}{|c|c|c|c|c||c|c|c|}
\hline
& \multirow{2}{*}{ $t \bar t W^+ j j$ }& \multirow{2}{*}{$t \bar t Z j j $} & \multirow{2}{*}{$t \bar t W^+ W^- $ }&\multirow{2}{*}{ $t \bar t  t \bar t$ } & \multicolumn{3}{|c|}{$ M_\pi ~(\mbox{TeV})$} \\ \cline{6-8}  & & & & & $0.9$ & $1.0$ & $1.2$ 
\\
\hline
no cut & 800 & 787 & 11.4 & 7.40 & 192 & 85.0  &  19.1
\\  
\hline
basic cuts (1--5) & 85.1  & 107  & 1.60 & 2.05 & 64.5 & 26.7 & 5.16
\\
\hline
$\begin{array}{c} p^{j1}_{T} > 100 ~\mbox{GeV},  p^{j2}_{T} > 50~\mbox{GeV}  
\\ ( p_{T}^{\ell^-} <10~\mbox{GeV}, \mbox{or}~ |\eta_{\ell^-} |> 2.5)  \end{array}$ &  
36.4 & 2.03 & 0.72 &  1.83 & 63.4 & 26.1 &  5.0
\\
\hline
$H_{T}>650~\mbox{GeV}$&28.1 & 1.36 & 0.51 & 1.68 & 63.2 & 26.0 & 4.99
\\
\hline 
$Acceptance$& $3.5 \%$ & $0.17 \%$ & $4.5 \%$ & $23 \%$ & $33 \%$ & $31\%$ & $26\%$
\\
\hline 
\end{tabular}
\end{center}
\caption{Number of events and final acceptance for the main SM backgrounds (not including fakes and charge mis-id) and for the signal from single and pair productions of $p ~ p  \to \bar t \bar t  \pi_6$, $t t  \pi_6^c$, $\pi_6  \pi_6^c$, $\pi_8 \pi_8 $ in an effective model with $a_R =1$. 
Numbers are given for an integrated luminosity of $\int {Ldt = 100} ~\mbox{fb}^{-1}$  at a $\sqrt s = 13$ TeV LHC.}
\label{cuttable}
\end{table}
\end{footnotesize}
\begin{footnotesize}
\setlength{\tabcolsep}{4pt}
\begin{table}[t!]
\begin{center}
\begin{tabular}{|c|c||c|c|c|c|c|c|c|c|}
\hline
& $\, M_\pi\, $ &  $0.9$ TeV   &  $1.0$ TeV &   $1.1$ TeV   &  $1.2$ TeV   &  $1.3$  TeV &  $1.4$ TeV  & $1.5$ TeV
\\
\cline{2-9}
\cline{2-9}
& $\pi_8 \pi_8$ &  $18.6$ & $7.60$  &  $3.06$ & $1.25$ & $0.55$ & $0.23$  & $0.10$
\\
\cline{2-9}
\hline
\multirow{3}{*}{$a_R = 1$} & $\pi_6 \pi_6^c $& $35.3$ &  $13.1$ & $4.99$  & $1.99$ &  $0.81$ & $ 0.32$ & $0.14$
\\  
\cline{2-9}
& $\pi_6 \bar t \bar t$ & $4.89$ &  $2.93$ & $1.75$ &  $1.01$ &  $0.60$ & $0.36$ & $0.22$
\\
\cline{2-9}
& $\pi_6^c t  t$ & $4.38$ &  $2.40$ & $1.35$ & $0.74$ & $0.42$ &  $0.25$ & $ 0.15$
\\
\hline
\hline
\multirow{3}{*}{$a_R = 2$} & $\pi_6 \pi_6^c $ &$24.2$ & $9.67$ &  $4.02$ & $1.76$  & $0.80$ &  $0.36$ & $ 0.18$ 
\\  
\cline{2-9}
& $\pi_6 \bar t \bar t$ & $16.8$ & $10.5$ &  $6.47$ & $4.02$ &  $2.62$ &  $1.72$ & $1.14$ 
\\
\cline{2-9}
& $\pi_6^c t  t$ & $15.1$ & $8.76$ &  $5.30$ & $3.38$ & $2.08$ & $1.35$ &  $0.94$
\\
\hline
\end{tabular}
\end{center}
\caption{Number of events for each channel with an integrated luminosity $\int {Ldt = 100} ~\mbox{fb}^{-1}$ at Run II after cuts. For the sextet, we used $a_R = 1$ (upper block) and $a_R = 2$ (lower block).}
\label{efficiency}
\end{table}
\end{footnotesize}

In the simulation we only included true SM backgrounds for the $2~ b $ + $2$ SSL + multi-jets signature, as fakes and charge misidentification can only reliably estimated with data based techniques. The list of the considered backgrounds, and their treatment, follows:
\begin{itemize}
\item  $t \bar t  W^\pm  + {\rm jets} $:  as we choose  two  positive  leptons as the signature, we are interested 
in the final states with  $ t \to  b \ell^+ v_\ell $, $ \bar t \to \bar b j j $ and $ W^+ \to \ell^+ v_{\ell} $. This  process represents the most sizable background for our signature.

\item  $t \bar t  Z  + {\rm jets} $:  this process yields a background for our signal if  $ t \to  b \ell^+ v_\ell $, $ \bar t \to \bar b j j $ and 
$ Z^+ \to \ell^+ \ell^- $. In addition to the basic cut,  
 we veto negatively charged electrons which have either $p_T > 10~\mbox{GeV}$ or $\eta_\ell < 2.5$.
This veto cut is crucial in our search strategy as it efficiently suppresses the $t \bar t Z$ background. While without this cut, the $t \bar t Z$ background efficiency is similar to the $t \bar t  W $ one (which is around $3.5\%$), the veto reduces the $t \bar t Z$ acceptance to negligible  $0.17 \%$.

\item  $t \bar t  W^+ W^- $:  this process yields a background if top and $W^+$ decay leptonically such that they generate the 2SSL.  
The anti-top and $W^-$ need to decay hadronically to produce additional jets to pass the event selection. The SM  cross section for this  process is small 
due to one additional massive gauge bosons present in the final state.

\item  four top production $t \bar t  t \bar t $:  this process corresponds to our signal final state, thus it is irreducible and  has high efficiency. However, the SM cross section of the process is small ($\sim$ two orders of magnitude lower than the cross section before cuts of the dominant background  $t \bar t  W^\pm  + {\rm jets} $), such that its final contribution to the background is subdominant.
\end{itemize}

\subsection*{Distinguishing sextets from octets}

Assuming that the LHC will collect enough statistics at Run II to see an excess in the 2SSL channel, one may ask the question if the signal is due to an octet or a sextet. The crucial difference between sextets and octets in the 2SSL search is that for sextets the SSLs originate from $\pi_6\rightarrow t t\rightarrow (l+\bar{\nu}b) (l+\bar{\nu}b )$, i.e. from the same resonance while for octets, each lepton originates from a top which result from the decay of two $\emph{different}$ $\pi_8$ resonances. This difference manifests itself very clearly in the angular distributions of the leptons and in the resonances' mass reconstruction, as we shown the following. 
 
 \bigskip
 
  Starting with the mass reconstruction, let us first consider an $l^+l^+$ signal arising from octet pair production. Reconstructing the $\pi_8$ mass requires to identify the correct combinations of $t_1\bar{t}_1\rightarrow(l^+\bar{\nu}b)_1(\bar{b}jj)_1$ and $t_2\bar{t}_2\rightarrow(l^+\bar{\nu}b)_2(\bar{b}jj)_2$. The reconstruction of semi-leptonic $t\bar{t}$ resonances is commonly applied, for example in $Z'$ or KK-gluon searches (although these searches only reconstruct one $t\bar{t}$ pair and not two as present in our case). For recently applied search strategies at ALTAS and CMS {\it c.f. e.g.} \cite{Aad:2015fna, Khachatryan:2015sma}. 

\bigskip

For the mass reconstruction of the sextet from the 2SSL channel with $l^+l^+$,  one can hope to reconstruct both the invariant mass of the leptonically decaying pair of tops (originating from $\pi_6 \to tt$ in our sample with $l^+ l^+$ pairs), and of the hadronically decaying ones (arising from $\pi_6^c \to \bar{t} \bar{t}$ in our sample).

The hadronically decaying resonance in $l^+l^+$  searches appears in the production channels of  $ p p  >  \pi_6  \pi_6^c $, $ t  t \pi_6^c $ from decays of the resonance $\pi_6^c$, while the other channels act as backgrounds. 
To reconstruct the 6-jet resonance we first  pick two jets whose combined invariant mass lies in the window of  $|m_{j_1j_2} - m_W| < 10~ \mbox{GeV}$ and identify them as a hadronic $W$.  Next, we use  another mass window criterion  $|m_{j_3 W}-m_t| < 10 ~\mbox{GeV}$  to  select a  third  jet or b-jet, which we pair with the 
reconstructed W bosons to an anti-top quark candidate.\footnote{Of course we cannot determine the charge of the resonance and only infer that it is an anti-top from the fact that we look at a 4-top signature with two positively charged leptons in the final state.} Applying the same procedure to the remaining jets and $b$-jets gives rise to another $W^-$ and anti-top reconstruction. Thus the four momentum of the $\pi_6^c$ particle can be fully  reconstructed from the hadronically decaying anti-tops as $p_{\pi_6^c}^\mu = p_{\bar t}^\mu + p_{\bar t}^\mu $. 

\begin{figure}[tb!]
\center
\includegraphics[scale=0.37]{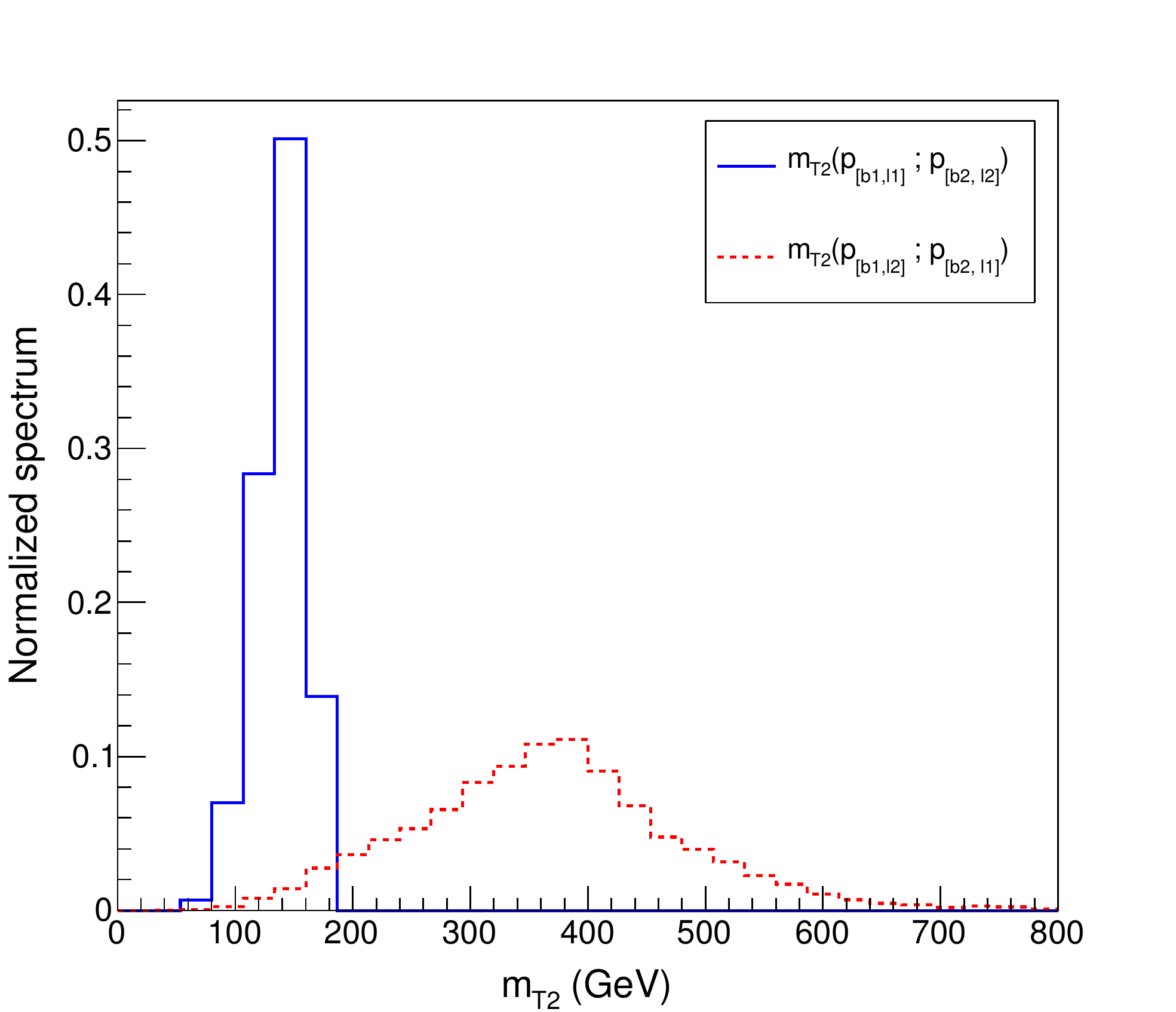}
\includegraphics[scale=0.37]{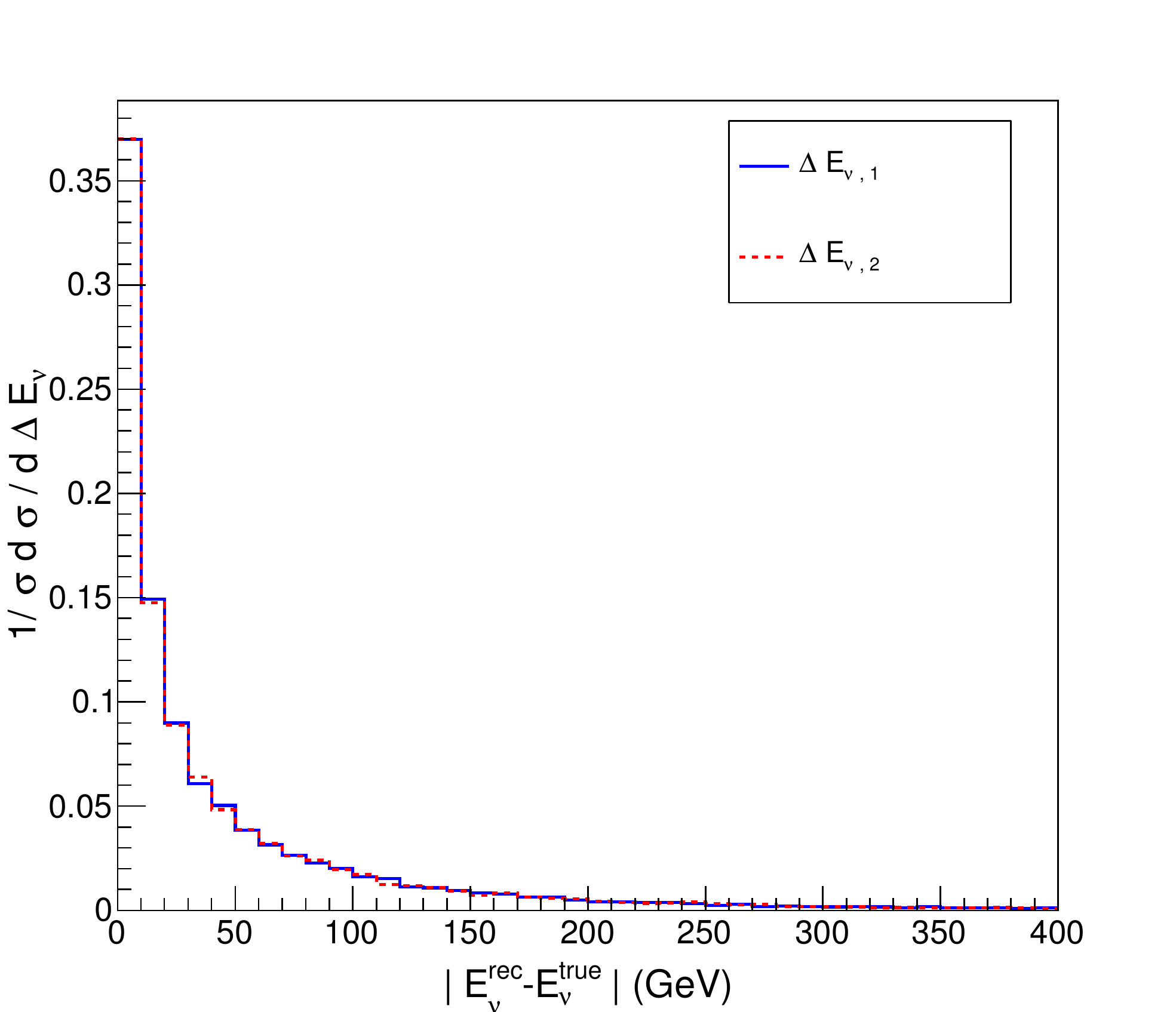}
\caption{Left panel: $m_{T2} $  distributions  for  two combinations of  $b$-$\ell^+$, the blue solid line corresponds to the correct combination and  the red dashed line corresponds  to the ``wrong'' one. The area  under  each distribution is normalized to be one. Right panel: distribution of $\Delta E_{\nu}$ between the true one generated by a Monte Carlo simulation and the reconstructed solution.}
\label{mt2pass}
\end{figure}

The positively charged di-leptons arise from the processes $ p p  >  \pi_6 \pi_6^c $ and $ \bar t  \bar t \pi_6$,  where $\pi_6$  decays to $2~ b$, $2$ SSL plus $\met$.\footnote{In the following, we discuss exclusively the $l^+l^+$ channel, but of course, for the $l^-l^-$ channel the analogous discussion holds when replacing particles by anti-particles.} The reconstruction procedure for the leptonically decaying top pair is much more complicated due to the missing energy from the two neutrinos.  In the  event selection,  we  demand 6 jets, two b-tagged jets, and  2 positive leptons. With the reconstruction of the two hadronic anti-tops, only two (possibly b-tagged) jets and the leptons remain which are the $b$'s and leptons from the two top decays in the underlying event. We face the problem to  correctly combine the $b$'s with the leptons to a  $(b_1, \ell_1)$ pair from one decay chain 
and the $(b_2, \ell_2)$ from the other one. We will demonstrate that the $m_{T2}$ variable can be used to identify the correct combination  in an very efficient way compared with other  possible variables,  e.g.  the invariant mass of the $(b_1, \ell_1)$ cluster.  The kinematic variable $m_{T2}$  is defined  in analogy to the transverse 
mass,  but extended to the case with two missing particles~\cite{Barr:2003rg, Cheng:2008hk}.   For  events with two missing particles  with mass  $\mu_N$   in  two  identical  decay chains, the quantity $m_{T2}^2$  is evaluated as  the  minimum  of  transverse mass square  over all  partitions of  the measured $\not{p}_T$, i.e. 
\begin{eqnarray}
\min_{\not{p}_T^1+ \not{p}_T^2= \not{p}_T} \left[ \max \{ m_T^2(  p_T^a, \, \mpt^1 ; \,\mu_N),\,
    m_T^2(  p_T^b, \,  \mpt^2; \,\mu_N)\} \right] ,
\end{eqnarray} 
where the transverse mass square $m_T^2$  is defined  as: 
\begin{eqnarray}
m_T^2 (  p_T^a, \, \mpt^1 ; \,\mu_N) =  m_a^2 + \mu_N^2 + 2 ( E_T^a \met^1 - p_T^a \cdot \not{p}_T^1)
\end{eqnarray} 
with $p_T^a$ being the transverse momenta of  a visible  cluster in one decay chain. In our case, we set  $p_T^a  =  (p_{b_1}^x + p_{\ell_1}^x, p_{b_1}^y + 
p_{\ell_1}^y)$ and $\mu_N = 0 $ since the neutrino mass is zero.  As shown in the left panel of Figure~\ref{mt2pass},  for the correct assignment  of the ($b$-$\ell^+$) cluster,  
$m_{T2}$  peaks around the top quark mass, while for the ``wrong'' combination its value is generally larger than 
the  top quark mass and peaks at a much higher energy scale. Thus in the simulation,  we identify the combination  $(b_1, \ell_1^+)$ as the one that gives the smallest $m_{T2}$.  The distributions plotted in Figure~\ref{mt2pass}  sum over  all the channels  with the individual  one weighted by the  cross 
section, respectively. In the case of  $M_\pi > 900 $ GeV,  for the channels $\pi_6 \pi_6^c$ and $\pi_6 \bar t \bar t$, the  $m_{T2}$ criterion  can achieve  a $99\%$ 
efficiency in order to get the correct combination. For the octet pair channel,  the efficiency can be as high as $94\%$;  while for the channel $\pi_6^c t t $, the 
efficiency is relatively lower, but still reaches  $87 \%$.

With the knowledge of  the correct combinations of $(b_1, \ell_1^+)$ and   $(b_2, \ell_2^+)$,   we are capable of reconstructing the sextet mass from the leptonic top 
decay by calculating  the four momenta of the two neutrinos. There are 6 unknown parameters  and a four-fold solution will  be achieved  by analytically  solving  2  
linear equations from missing transverse momenta and  4 nonlinear equations from the on-shell conditions of  $W$ mass and top quark mass~\cite{Sonnenschein:2006ud, Bai:2008sk}.  The fourth-degree  polynomial  for one neutrino energy can be expressed as:
\beq
c_0 + c_1 E_{v}  + c_2 E_{v}^2 + c_3 E_{v}^3 + c_4 E_{v}^4 = 0
\eeq
where those coefficients $c_i$ are complicated functions  of  $m_W$, $m_t$  and momentum of  leptons and  bottom quarks.  Solving this equation will lead to  
the possibility of  $0$, $2$  or $4$ real solutions and in this process, we shall apply  an additional cut to require  $E_\nu < 1500 $ GeV.  Events with zero  
solutions will be discarded.  Although the solutions satisfying the  minimum kinematic constraints can all be generated, neutrinos in a certain energy 
range are more accessible at the LHC. Therefore we can  resolve  this ambiguity  by  choosing  the solution with the smallest $|p_x|$  since the 
process  being  investigated  does not carry large missing energy. In the right panel of Figure ~\ref{mt2pass}, we show the statistical  distribution for  the  difference in energy
$\Delta E_{\nu} $ of the  reconstructed neutrino and  the MC generated one.   The possibility for  either one of  energy differences satisfying 
$\Delta E_{\nu, i} < 50 $ GeV  with $i = 1, 2$  reaches  $80 \%$,  which validates  our  choice  for the neutrinos with the smallest absolute value for the 
x-axis momentum. The  mass of sextet is  thus  determined  by  the invariant mass of  the reconstructed $t t$ pair. 

\begin{figure}[tb!]
\center
\includegraphics[scale=0.4]{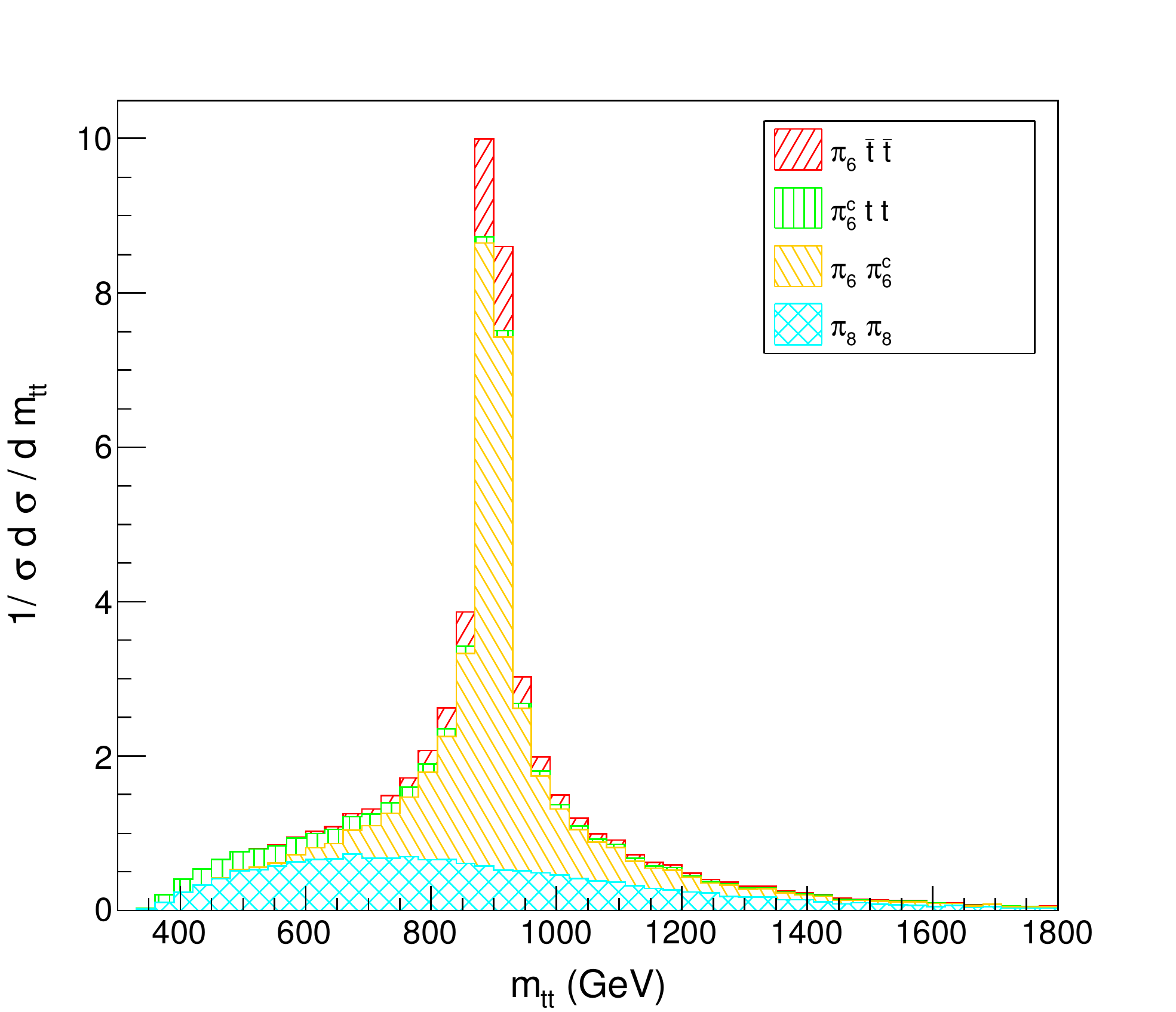}
\caption{Staggered plot of the invariant mass  $m_{t t} $ distribution from dilepton decay for the case of $M_\pi = 900$ GeV and  $a_R = 1$ for the channels $\pi_6\bar{t}\bar{t}, \pi_6^c t t, \pi_6^c\pi_6$, and $\pi_8\pi_8$. The peak is  reconstructed after passing  the basic cuts and  the b-tag efficiency is included.}
\label{mass_pi6}
\end{figure}

Figure~\ref{mass_pi6} shows reconstructed invariant mass of the $tt$ pair from the leptonic decay. We see that the channels with a $\pi_6$ in the final state generate a clear peak around the true mass (which will be smeared by detector effects), while processes with non-resonant $tt$ pair, originating from the octet pair production or the tops associated with a single $\pi_6^c$, produce a continuum background.
Setting a cut on the reconstructed invariant mass can thus allow to isolate the contribution of the $\pi_6 \to tt$ decays and reveal the presence of a sextet in the data. 

 A further invariant mass can be reconstructed from the hadronic $\bar{t}\bar{t}$ pair.\footnote{This procedure has been carried out in Ref.~\cite{Chen:2008hh}.} For $m_{\bar{t}\bar{t}}$ , the processes $\pi_6^c\pi_6$ and $\pi_6^c t t$ yield a contribution which is peaked around $M_{\pi}$ while $\pi_8\pi_8$ and $\pi_6\bar{t}\bar{t}$ contribute with flat distributions. Using both mass invariants in parallel thus even in principle allows to directly determine the coupling $a_R$, as in $\pi_6^c\pi_6$ events, both the leptonic and the hadronic invariant masses reproduce $M_\pi$ simultaneously, while $\pi_6 \bar{t} \bar{t}$ reconstructs leptonically but not hadronically and vice versa for $\pi_6^c t t$.\footnote{An indirect measurement of $a_R$ could also be obtained by comparing the mismatch of the number of events with a reconstructed $tt$ resonance to the prediction of such events by QCD pair production of such events for the mass determined from the measurement.} Here, we do not quantify the separation power of the direct measurement of $a_R$ as its significance depends on the backgrounds, and in our analysis we can only include the true SM background, while the background from fake 2SSL events and 2SSL events which arise from charge mis-identification needs to be extracted from actual data. We just wish to point out that reconstructing the $\pi_6$ mass in two independent ways in principle allows for a direct measurement of $a_R$.

\bigskip

\begin{figure}[tb!]
\center
\includegraphics[scale=0.37]{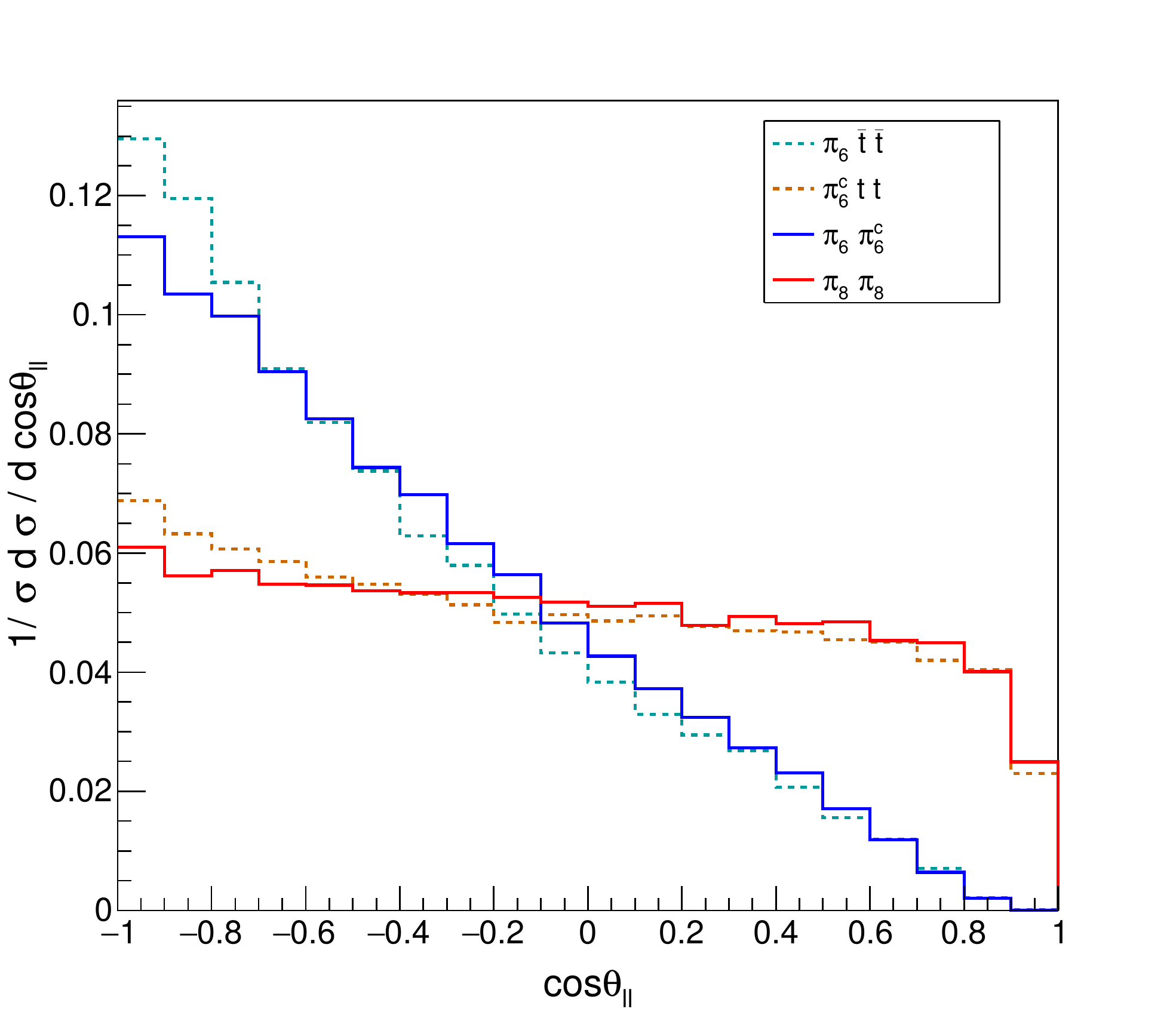}
\includegraphics[scale=0.37]{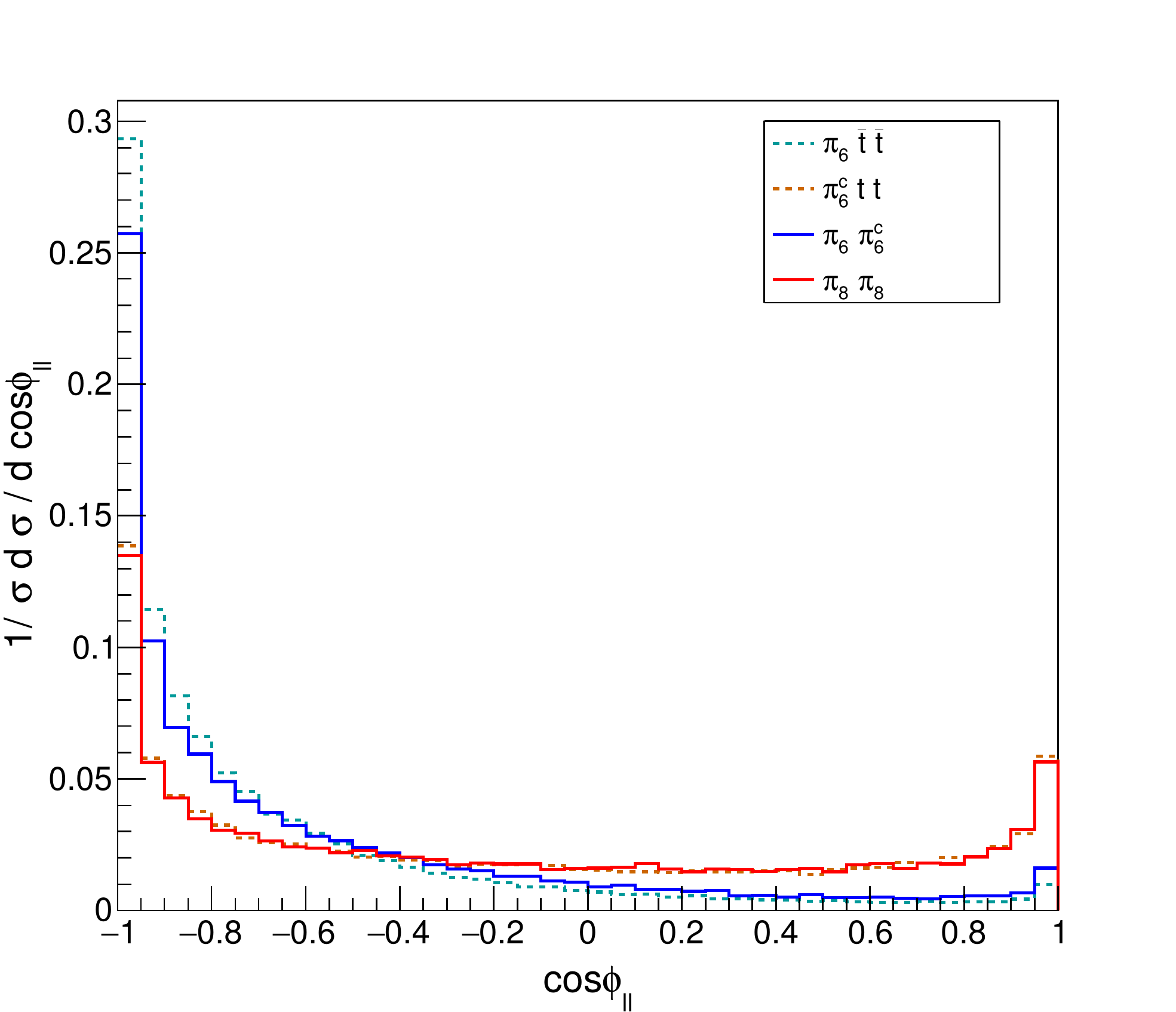}
\caption{Left panel: dilepton opening angle  distribution in the laboratory frame of the different $4t$ channels (and $M_\pi = 900~\mbox{GeV}$)  after passing the basic cuts.\newline
 Right panel: dileptonic azimuthal angle difference distribution  in the laboratory frame for the same channels and $M_\pi = 900~\mbox{GeV}$, after passing the basic cuts.}
\label{cosphi2l}
\end{figure}

Also, the angular distributions of the leptons carry important information about their origin. One simple way to differentiate the sextet from the octet without boosting the lepton momenta is to use the opening angle between the 2SSL in the detector 
frame, which is defined as:
\beq
\cos \theta_{\ell_1^+ \ell_2^+} =  \frac{ 
\vec{p}_{\ell_1} \cdot \vec{p}_{\ell_2} }{\left| \vec{p}_{\ell_1} \right| \left| \vec{p}_{\ell_2} \right|}\,.
\eeq
In the case of a 2SSL from a $\pi_6$ decay, the opening angle tends to be large for two reasons. First, as $M_\pi\gtrsim 800~\mbox{GeV}$ from Run I constraints already, the $\pi_6$ decays into two  boosted tops which have a large opening angle as the $\pi_6$ itself will be dominantly produced at low boost (in single as well as in pair production). As the tops are highly boosted, their decay products (and in particular the leptons) are approximately collinear to the boost direction, and therefore also the 2SSL tend to have a large opening angle. Second, according to the lepton correlation with the top spin in the top decay, the lepton tends to move in the same direction  as  the top quark momentum  for a pure $t_R$ case. Thus, in the case  that  the two leptons are coming from one  sextet resonance which in our model couples purely to right-handed tops,  they are more likely to be back-to-back thus giving a maximal opening angle.  
For the other cases  where the  dilepton  comes from unpolarised top quarks and/or from the decay of two different resonances,  the  opening angle distribution  will be  more spherical. The differential distribution $1/ \sigma \cdot d \sigma /  d \cos \theta_{\ell_1^+ \ell_2^+}$ for a mass $M_\pi = 900~\mbox{GeV}$, shown in the left panel of Figure~\ref{cosphi2l}, illustrates this fact. For larger $M_\pi$, the 2SSL opening angle increases even further for events with a $\pi_6$, as the tops arising from its decay are more boosted.

\bigskip

Alternatively (or in addition), one can use the difference in azimuthal angles $\phi_{\ell_1^+ \ell_2^+}$ with $\vec{p}_{\ell_1}$ and $\vec{p}_{\ell_2}$  in the laboratory frame to 
trace the spin correlation between the two leptons. The advantage of this variable is that  this information can be easily obtained at the LHC and 
at the same time it is sensitive to the polarisation of the top quarks. In analogy to the $t \bar t $ pair decay \cite{Mahlon, Bernreuther},  we  define  the  
azimuthal angle correlation  in the scenario of 2SSL to be:
\beq
\cos \phi_{\ell_1^+ \ell_2^+} =  \frac{ 
\vec{p}_{\ell_1} \cdot \vec{p}_{\ell_2} - p_{\ell_1}^{z}  p_{\ell_2}^{z}  }{\left| \vec{p}_{\ell_1} \right| \left| \vec{p}_{\ell_2} \right|}
\eeq
In the right panel of Figure~\ref{cosphi2l},  we plot the distribution of $1/ \sigma \cdot d \sigma /  d \cos \phi_{\ell_1^+ \ell_2^+}$  for  each channel, again for $M_\pi = 900~\mbox{GeV}$. In  both cases -- whether the 2SSL arises from a $\pi_6$ decay or not --   the  azimuthal 
angle difference  peaks towards the  back-to-back configuration, however the azimuthal angle asymmetry in the sextet channel is  much more pronounced than in the octet  
channel.  For  the octet  case,  it is the momentum  conservation which forces  the  2 SSL  coming from two  different resonances to be back to back in the moving direction,  while  for the sextet case the two leptons from one resonance will also fly in the opposite direction due to the helicity conservation.

\begin{figure}[tb!]
\center
\includegraphics[scale=0.4]{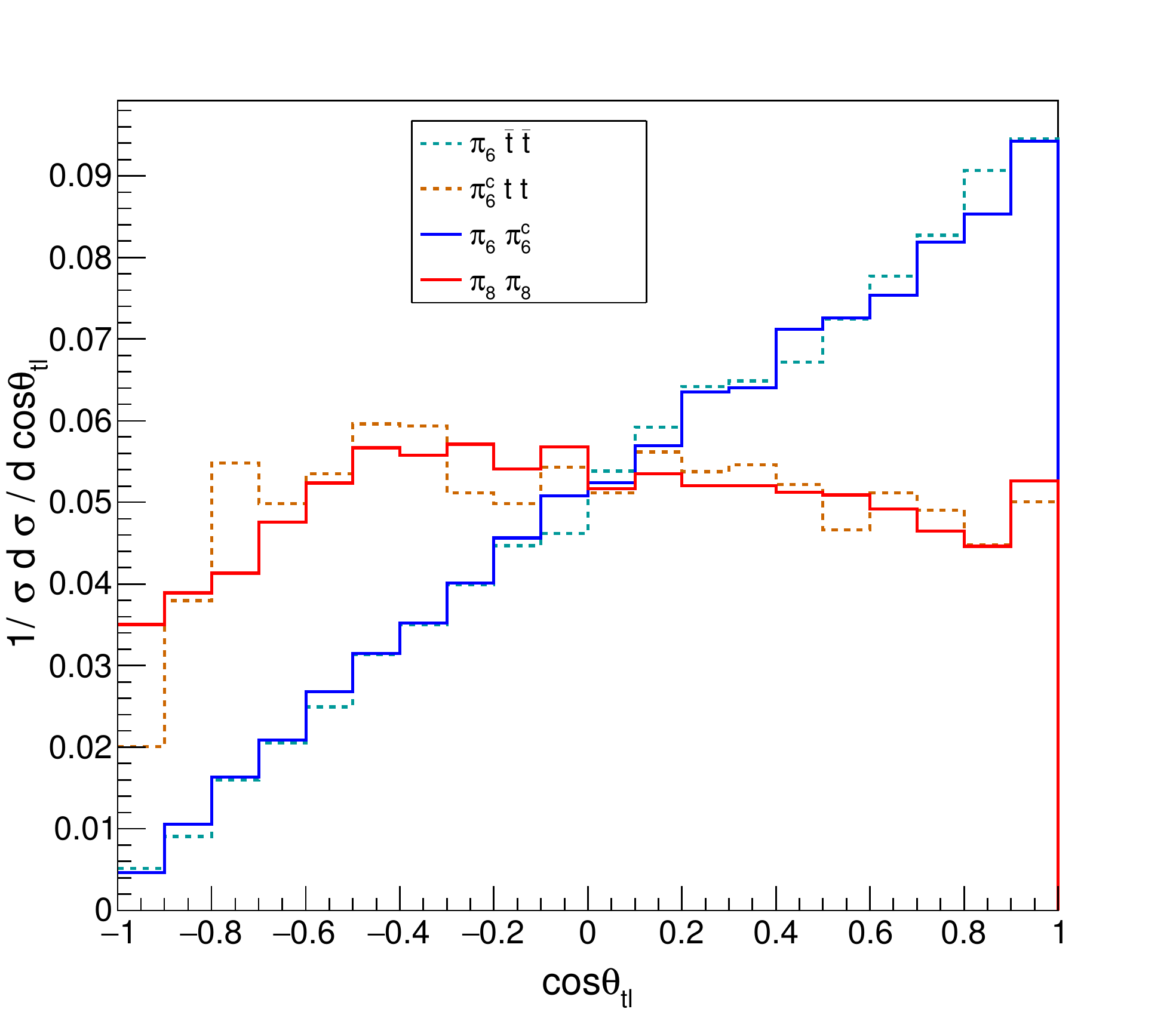}
\caption{Leptonic angular distribution relative to the top quark momentum  for  the events with top quark reconstructed and  passing the basic cut. Note that we also impose a mass window selection rule $ |m_{tt}-900 ~\mbox{GeV}| < 100~\mbox{GeV} $. }
\label{costhetatl}
\end{figure}

\bigskip

Apart from considering angular distributions in the laboratory frame, we are also considering angular distributions in the rest-frame of the tops in order to directly access the chirality of the tops produced. These distributions naively appear to be harder to access as the tops are highly boosted such that the angular resolution in the rest frame of the top are subject to large uncertainties but we nevertheless find that these distributions yield some sensitivity. In the case under study, the tops originating from the decays of the sextet are dominantly right-handed, while the tops from the octet and the ones associated to the singly produced sextet have mixed chirality.
The chiral structure of the top quarks can be identified by using the leptonic angular distribution in the rest frame of the top quark: we can define $\theta_{t \ell}$ to  be  the  angle  between  the lepton  momentum  and  the chosen spin axis for the top quark, i.e.  its spatial  momentum  in the rest  frame of  $t t$ pair.  The 
differential decay rate is simply described by~\cite{kuhn,Parke,Stelzer}:
\beq
\frac{1}{\sigma}\frac{d \sigma } { d \cos \theta_{t \ell^+} } = \frac{1}{2} (1 + a \cos \theta_{t \ell^+} ),
\eeq
with $a =1 $ for a  pure $t_R$ and $a =-1 $ for a pure $t_L$.   In  Figure~\ref{costhetatl}, we display the $\cos\theta_{t \ell^+}$ distributions for each channel for $M_\pi = 900 ~\mbox{GeV}$, where 
we boost the $p_{\ell^+}$ of the positive lepton into the rest frame of top quark and  also boost the $p_t$ into the rest frame of the $t t$ pair cluster.  For  the cases  
$ \pi_6  \bar t  \bar t $, $  \pi_6 \pi_6^c  $, with the positive lepton $\ell^+ $  as  a  daughter particle  from  the scalar $\pi_6$, the events  show  a  clear distribution 
of   $(1+ \cos \theta_{t \ell^+})$, which verifies that  a sextet scalar only couples to  right handed tops.  On the other hand, for the other cases $ \pi_6^c   t  t $ and $  \pi_8 \pi_8 $, 
the distributions are substantially flatter.

\section{Conclusions}
\label{sec:conclusion}

Fundamental composite electroweak dynamics (FCD) is an alternative way to build models of the electroweak interactions where the Higgs boson arises as a composite state of a confining dynamics with fermionic components.
This approach allows to identify the light resonances one may expect in the low energy theory, thus providing a precious guidance in the construction of effective models of a composite Higgs.
Requiring a fermionic UV completion leads to the immediate conclusion that one may well expect more scalar mesons than a Higgs-like singlet and the eaten Goldstone bosons.
In particular, the requirement that the theory generates fermionic bound states (baryons) that carry colour generically imply the presence of coloured mesons in the spectrum, whose mass is linked to the mass of the baryons themselves. The baryons play the role of top partners in the partial compositeness paradigm by mixing linearly with the top (and bottom) to provide them a mass.

We focused on the scenario where the coloured scalars are lighter that the coloured baryons: in this case, they dominantly couple to top quarks via the mixing dictated by partial compositeness. We considered a charged sextet coupling to a di-top state, and real octet coupling to $t\bar{t}$. We then studied the phenomenology at the LHC, both Run I and II, by using of a simplified Lagrangian approach. Gauge invariance under the EW gauge group requires that the sextet couples dominantly to right handed tops with an unsuppressed (order 1) coupling. We thus considered both pair production of sextet and octet, and the single production of the sextet in association to an anti-top pair.
All production channels give rise to 4-top final states, which are very well constrained by searches focusing on two same sign leptons, and one lepton plus jets.
The searches at Run I on the full dataset provide a lower bound on the mass of the new scalars ranging from 800 GeV to 1.1 TeV.
We then investigated the perspectives for Run II, and in particular  the possibility of distinguishing the presence of a sextet in the same sign lepton search from the presence of an octet once a significant excess is detected.
Same sign leptons are promising as they originate from the decay of a single resonance $\pi_6 \to t t \to b b l^+ l^+ \met$, thus giving a handle on the kinematics of the resonance. We illustrated three methods apt to single out the sextet from the signal.
The first possibility is to reconstruct the resonance by fully reconstructing the leptonically decaying tops: we demonstrated that this can be done with very high efficiency by a simulation at particle level. Detector effects will smear the peak, however a distinction from the smooth distribution provided by the octet is possible.
The other two methods rely on the angular distributions of the leptons, and rely on the fact that the sextet decays to two right handed tops (and not a $t\bar{t}$ pair).
One method we presented relies on studying angular distributions between the two leptons in the laboratory frame, which are produced more back-to-back from the sextet decays then from other channels.
The other method consists on studying the chirality of the leptonically decaying tops, showing a marked right-handed polarisation in the signal events from the sextet.

The set up we studied may be rather general, however we present an explicit model where such states are predicted. The model is based on a coset SU(4)/Sp(4) in the EW sector, giving rise to a Higgs-like state plus a singlet.
QCD colour interactions are included by adding a second species of fundamental fermions, thus adding a global symmetry SU(6)/SO(6) which generates a neutral octet and a charged sextet as pseudo-Nambu-Goldstone bosons.
We studied how the masses can be generated, showing that they can be naturally degenerate at the TeV scale. We further investigated the couplings to fermions arising via the partial compositeness couplings.
Besides the aforementioned couplings to tops, the coloured mesons also couple to the top partners.
Depending on the detailed spectrum, one may therefore expect decays of the top partners into the coloured scalars, thus weakening the bounds on vector-like top partners from present searches while at the same time boosting the production cross section of coloured scalars further. Another interesting possibility would be that the scalars (if heavier than the top partners) may decay into top partners, thus providing an additional production mechanism for them.

\section*{Acknowledgements}
This work was supported by the National Research Foundation of Korea (NRF) grant funded by the Korea government (MEST) (No. 2012R1A2A2A01045722), the International Research \& Development Program of the National Research Foundation of Korea (NRF) funded by the Ministry of Science, ICT \& Future Planning (Grant number: 2015K1A3A1A21000234), and by the Basic Science Research Program through the National Research Foundation of Korea (NRF) funded by the ministry of Education, Science and 
Technology (No. 2013R1A1A1062597). We thank the France-Korea Particle Physics Lab (FKPPL) for partial support. AD is partially supported by Institut 
Universitaire de France. AP acknowledges IBS Korea for support under system code IBS-R017-D1-2015-a00. We also acknowledge partial support from the Labex-LIO (Lyon Institute of Origins) under grant ANR-10-LABX-66 and FRAMA 
(FR3127, F\'ed\'eration de Recherche ``Andr\'e Marie Amp\`ere").

\appendix

\section{Pre-Yukawa coupling structures} \label{app:1}

The expansions for the 4 pre-Yukawa couplings read:
\beq
y_{5R} f\, \xi_R U_4^{(R)} \psi_5 &=& \frac{y_{5R} f}{\sqrt{2}} \left[ t_R^c (T + X_{2/3}) \left( \sin \epsilon + \cos \epsilon \frac{h}{\sqrt{2}} \right) + i t_R^c \tilde{T}_5 \cos \epsilon\, \frac{\eta}{f} + \dots \right]\,, \\
y_{5R} f\, \xi_R U_4^{(R)} \eta_5 &=& - \frac{y_{5R} f}{\sqrt{2}} \left[ t_R^c (T^c + X_{2/3}^c) \sin \epsilon + \dots \right]\,;
\eeq
\beq
y_{1R} f\, \xi_R U_4^{(R)} \psi_1 &=& y_{1R} f \left[ t_R^c \tilde{T}_1 \left( \cos \epsilon - \sin \epsilon \frac{h}{\sqrt{2}} \right) + \dots \right]\,, \\
y_{1R} f\, \xi_R U_4^{(R)} \eta_1 &=& y_{1R} f \left[ t_R^c \tilde{T}_1^c \cos \epsilon + \dots \right]\,;
\eeq
\beq
y_{5L} f\, \xi_L U_4^{(R)} \eta_5 &=& y_{5L} f \left[  B^c b_L +  T^c t_L \cos^2 \frac{\epsilon}{2} -  X^c_{2/3} t_L \sin^2 \frac{\epsilon}{2} + \right. \nonumber \\
& & \left. - \frac{1}{2} (T^c + X^c_{2/3}) t_L \sin \epsilon\, \frac{h}{\sqrt{2}} + i \frac{1}{2} \tilde{T}_5^c t_L \sin \epsilon \frac{\eta}{f} + \dots \right]\,, \\
y_{5L} f\, \xi_L U_4^{(R)} \psi_5 &=& y_{5L} f \left[ X_{5/3} b_L + T t_L \sin^2 \frac{\epsilon}{2} - X_{2/3} t_L \cos^2 \frac{\epsilon}{2} + \dots \right]\,;
\eeq
\beq
y_{1L} f\, \xi_R U_4^{(R)} \eta_1 &=& \frac{y_{1L} f}{\sqrt{2}} \left[  \tilde{T}_1^c t_L \left( \sin \epsilon + \cos \epsilon \frac{h}{\sqrt{2}} \right) + \dots \right]\,, \\
y_{1L} f\, \xi_R U_4^{(R)} \psi_1 &=& \frac{y_{1L} f}{\sqrt{2}} \left[ \tilde{T}_1 t_L \sin \epsilon + \dots \right]\,.
\eeq

The couplings of the octet can be obtained from \refeq{eq:leftpreY} and \refeq{eq:rightpreY}, and are proportional to the mass matrix (with $M_{5,1} = 0$):
\beq
g_{\pi_8 t_L t^c_R} =- i \frac{f}{\sqrt{2} f_6} \left( \begin{array}{ccccc}
0 & y_{5L} \cos^2 \frac{\epsilon}{2} & - y_{5L} \sin^2 \frac{\epsilon}{2} & \frac{y_{1L}}{\sqrt{2}} \sin \epsilon & 0 \\
\frac{y_{5R} }{\sqrt{2}} \sin \epsilon & 0 & 0 & 0 & 0 \\
\frac{y_{5R} }{\sqrt{2}} \sin \epsilon & 0 & 0 & 0 & 0 \\
y_{1R}  \cos \epsilon & 0 & 0 & 0 & 0 \\
0 & 0 & 0 & 0 & 0 
\end{array} \right)\,,
\eeq
in the gauge eigenstate basis.
The couplings of the sextet can be written as:
\beq
g_{\pi_6 t^c_R t^c_R} &=& - i \frac{f}{2 \sqrt{2} f_6} \left( \begin{array}{ccccc}
0 & - \frac{y_{5R} }{\sqrt{2}} \sin \epsilon & - \frac{y_{5R} }{\sqrt{2}} \sin \epsilon & y_{1R}  \cos \epsilon & 0 \\
- \frac{y_{5R} }{\sqrt{2}} \sin \epsilon & 0 & 0 & 0 & 0 \\
- \frac{y_{5R} }{\sqrt{2}} \sin \epsilon & 0 & 0 & 0 & 0 \\
y_{1R}  \cos \epsilon & 0 & 0 & 0 & 0 \\
0 & 0 & 0 & 0 & 0 
\end{array} \right)\,; \\
g_{\pi_6^c t_L t_L} &=& - i \frac{f}{2 \sqrt{2} f_6} \left( \begin{array}{ccccc}
0 & y_{5L} \sin^2 \frac{\epsilon}{2} & - y_{5L} \cos^2 \frac{\epsilon}{2} & \frac{y_{1L}}{\sqrt{2}} \sin \epsilon & 0 \\
y_{5L} \sin^2 \frac{\epsilon}{2} & 0 & 0 & 0 & 0 \\
 - y_{5L} \cos^2 \frac{\epsilon}{2}  & 0 & 0 & 0 & 0 \\
\frac{y_{1L}}{\sqrt{2}} \sin \epsilon & 0 & 0 & 0 & 0 \\
0 & 0 & 0 & 0 & 0 
\end{array} \right)\,.
\eeq
For the bottom, in the basis $\{ b, B\}$,
\beq
M_{\rm bottom} = \left( \begin{array}{cc}
0 & y_{5L} f \\
0 & M_5
\end{array} \right)\,,
\eeq
and the coupling of the octet and sextet reads:
\beq
g_{\pi_8 b_L b_R^c} = - i \frac{f}{\sqrt{2} f_6}\left( \begin{array}{cc}
0 & y_{5L} \\
0 & 0
\end{array} \right)\,, \quad g_{\pi_6^c b_L X_{5/3}} = - i \frac{f}{\sqrt{2} f_6} y_{5L}\,.
\eeq


\end{document}